\documentclass[%
 aip,
 amsmath,amssymb,
preprint,%
]{revtex4-1}

\DeclareSymbolFont{matha}{OML}{txmi}{m}{it}% txfonts
\DeclareMathSymbol{\varv}{\mathord}{matha}{118}
\usepackage[normalem]{ulem}

\usepackage{graphicx}% Include figure files
\usepackage{float}
\usepackage{dcolumn}% Align table columns on decimal point
\usepackage{bm}% bold math
\usepackage[utf8]{inputenc}
\usepackage[T1]{fontenc}
\usepackage{mathptmx}
\usepackage{etoolbox}
\usepackage{soul}
\usepackage[T1]{fontenc}
\usepackage[utf8]{inputenc}
\usepackage[absolute,overlay]{textpos}
\usepackage{hyperref}
\usepackage{graphicx}
\usepackage{dcolumn}
\usepackage{xcolor}
\usepackage{siunitx}
\usepackage[all]{nowidow}
\usepackage{subfig}

\usepackage{commath} % \abs{} -> |a|; \norm{] -> ||a||
\usepackage{braket}  % \braket{} -> <a>
\usepackage{lipsum} 
\usepackage{import}
\usepackage{color}
\newcommand*{\citen}{}% generate error, if `\citen` is already in use
\DeclareRobustCommand*{\citen}[1]{%
  \begingroup
    \romannumeral-`\x % remove space at the beginning of \setcitestyle
    \setcitestyle{numbers}%
    \cite{#1}%
  \endgroup
}

\makeatletter
\def\@email#1#2{%
 \endgroup
 \patchcmd{\titleblock@produce}
  {\frontmatter@RRAPformat}
  {\frontmatter@RRAPformat{\produce@RRAP{*#1\href{mailto:#2}{#2}}}\frontmatter@RRAPformat}
  {}{}
}%
\makeatother

\begin{document}

%\preprint{AIP/123-QED}

\title[]{Phase Transitions of Oscillating Droplets on Horizontally Vibrating Substrates  }
% Force line breaks with \\
\author{King L. Ng*}
\email{klng@ifpan.edu.pl}
\affiliation{Institute of Physics, Polish Academy of Sciences, Al. Lotnik\'ow 32/46, 02-668 Warsaw, Poland}
\author{Lu\'is H. Carnevale}
\affiliation{Institute of Physics, Polish Academy of Sciences, Al. Lotnik\'ow 32/46, 02-668 Warsaw, Poland}
\author{Michał Klamka}
\affiliation{Institute of Aeronautics and Applied Mechanics, Warsaw University of Technology, Nowowiejska 24, 00-665 Warsaw, Poland}
\author{Piotr Deuar}%
% \email{deuar@ifpan.edu.pl}
\affiliation{Institute of Physics, Polish Academy of Sciences, Al. Lotnik\'ow 32/46, 02-668 Warsaw, Poland}
\author{Tomasz Bobinski}
% \homepage{http://www.Second.institution.edu/~Charlie.Author.}%
\affiliation{%
Institute of Aeronautics and Applied Mechanics, Warsaw University of Technology, Nowowiejska 24, 00-665 Warsaw, Poland%\\This line break forced% with \\
}%
\author{Panagiotis E. Theodorakis}%
 %\email{panos@ifpan.edu.pl}
\affiliation{Institute of Physics, Polish Academy of Sciences, Al. Lotnik\'ow 32/46, 02-668 Warsaw, Poland}

\date{\today}% It is always \today, today,

\begin{abstract}
Droplet deformations caused by substrate vibrations are
ubiquitous in nature and highly relevant for applications 
such as microreactors and single-cell sorting. 
The vibrations can induce droplet oscillations, a fundamental process that 
requires an in-depth understanding. Here, we report on extensive
many-body dissipative particle dynamics simulations carried out to
study the oscillations of droplets of different
liquids on horizontally vibrating substrates,
covering a wide range of vibration frequencies and amplitudes 
as well as substrate wettability. We categorize the phases
observed for different parameter sets based on the capillary number
and identify the transitions between the observed oscillation phases, 
which are characterized by means of suitable parameters, such as the angular
momentum and vorticity of the droplet. The instability growth rate
for oscillation phase II, which leads to highly asymmetric oscillations 
and eventual droplet breakup, is also determined.
Finally, we characterize the state of the droplet for the various 
scenarios by means of the particle--particle and particle--substrate contacts. 
We find a steady-state scenario for phase I,
metastable breathing modes for phase II, and an out-of-equilibrium
state for phase III. Thus, we anticipate that this study provides much needed
insights into a fundamental phenomenon in nature with significant relevance for applications.
\end{abstract}

\maketitle

\section{Introduction}
Deformed droplets are ubiquitous in industrial applications such
as ink-jet printing,\cite{Hoath2016} spray cooling,\cite{Yin2022} 
and combustion,\cite{Kummitha2024} as well as in 
lab-on-a-chip applications.\cite{Dkhar2023,Wu2025}
The deformations can be the result of substrate (or wall) vibrations
in contact with the droplets and can impact the performance
of technologies at a fundamental level.\cite{Timonen2013,Dong2006,Manor2011}
For example, they can favorably be exploited in applications
such as microfluidics,\cite{Mugele2006,Daniel2005,Chang2013} directional
motion,\cite{Dong2017,Abubakar2022,Lu2025} or in an
oscillating droplet tribometer,\cite{Junaid2022} 
but may be undesirable in situations where 
droplet breakup and the formation of 
satellite droplets\cite{carnevale2023} must be avoided, as
in the case of ink-jet printing.\cite{Hoath2016}
Moreover, high-frequency oscillations attributed to rotation
effects hold importance for applications in drug delivery,\cite{Kang2020} mass-transfer enhancement in microreactors,\cite{Zhao2007} and high-throughput single-cell sorting.\cite{Zhang2024b}

The focus is therefore placed here on droplet oscillations on vibrating,
flat, solid substrates.\cite{Rodot1979} The background case of free droplets
has been the subject of research for many 
decades,\cite{Daniel2005,book:chandrasekhar,Mettu2008,Lyubimov2006}
initially by Kelvin\cite{Kelvin1890} and Lamb\cite{Lamb1881}
more than a century ago,
and later by Rayleigh.\cite{Rayleigh1945}
These studies have provided theoretical descriptions of
surface oscillation modes, categorized according to the dominant restoring
force for the oscillations, such as surface tension or gravity.
In the case of constrained drops, oscillations have been
considered only in the past few decades\cite{Rodot1979} and
can generally be classified as axisymmetric or non-axisymmetric surface
oscillations, depending on whether the substrate vibrates
vertically (axisymmetric oscillation)\cite{Chang2013} or horizontally.\cite{Milne2014} 
One should note, however, that the Rayleigh half-drop solutions
to the sessile drop governing equations, which, along with their 
frequencies, make up the Rayleigh spectrum, also contain spherical harmonics
that include shapes breaking the axisymmetry of the Rayleigh
drops.\cite{Chang2013,Lamb1932, book:chandrasekhar} 
In particular, non-axisymmetric sectorial oscillations of water drops have been excited through parametric resonance by using acoustic levitation and an active modulation
method.\cite{Shen2010} 
However, describing the droplet oscillations and extracting information
on the emerging velocity profiles from experiments
are in general challenging. Typically, one must assume that 
patterns remain periodic, substrates are ideally homogeneous,
and evaporation takes place beyond a time spanning many
oscillation periods. 
For these and other reasons, even the description of the natural modes remains elusive despite numerous studies.\cite{Strani_Sabetta_1984,Chiba2012,Lyubimov2006,Noblin2004,Sharp2011,Sharp2012,Smithwick1989}
On the theoretical side, various concepts have been utilized
to describe droplet oscillations, such as that of coinciding
frequency with one of the natural frequencies,\cite{Kalmar-Nagy2011,Chen2016,Rahimzadeh2019,Mettu2008,Celestini2006,Deepu2014,Lin2018,Li2024}
or models such as a nonlinear mass-spring-damper known as the Duffing oscillator,\cite{Kalmar-Nagy2011}
which has been used to study axisymmetric modes of sessile droplets
subjected to vertical vibrations.\cite{Deepu2014} 
As in the case of experiments, theoretical work also comes with limitations, predominantly adopting various assumptions, such as
those regarding the type of restoring force and neglecting the complex nature of oscillatory phenomena. For example, oscillations can be influenced by various factors such as internal
flow within the droplets,\cite{Ng2025} liquid--vapor 
coexistence, evaporation,\cite{Theodorakis_MCVOF_2021} 
contact-line dynamics,\cite{Karim2022} 
physical pinning,\cite{Theodoraksi2021_pinning} and others.
Unfortunately, most of these phenomena
continue to pose challenges for theoretical approaches, 
despite sustained efforts.\cite{Rahimzadeh2019}

Considering the above limitations of both theory and experiment,
we have recently embarked on investigating droplet oscillations
on horizontally vibrating solid substrates by means 
of computer simulations as an alternative approach. Specifically, we have employed many-body dissipative particle dynamics (MDPD),\cite{Ng2025}
a suitable simulation method for reaching large droplet
sizes with molecular resolution while retaining the capability to
describe relevant droplet properties such as flow fields
and contact angles in detail, 
without utilizing any \textit{a priori} assumptions like
a contact angle model. Moreover, MDPD allows the study of 
various types of liquids, including complex fluids 
(e.g., fluids with surfactants\cite{Carnevale2024}), at a
computational cost much lower than that of traditional 
molecular dynamics.\cite{Carnevale2024_MDPD_MARTINI,Kramarz2025}
A wide spectrum of oscillation frequencies and amplitudes for 
various substrate-wettability cases can be explored in the simulations, 
without posing significant challenges often faced in experiments,\cite{Gilewicz_Master_Thesis}
where, for example, one needs to consider limitations 
concerning the speed of the motor that controls the substrate vibrations,
the choice of the equilibrium contact angle,
and ensuring homogeneous substrate properties.

\begin{figure}[htb!]
    \subfloat[\centering ]{{\includegraphics[width=0.33\columnwidth,trim=4.5cm 12cm 6cm 8.5cm,clip]{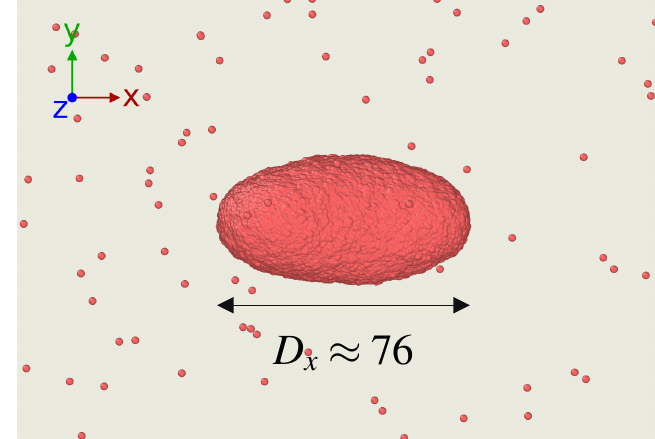} }}%
    %\quad
    \subfloat[\centering ]{{\includegraphics[width=0.33\columnwidth,trim=4.5cm 12cm 6cm 8.5cm,clip]{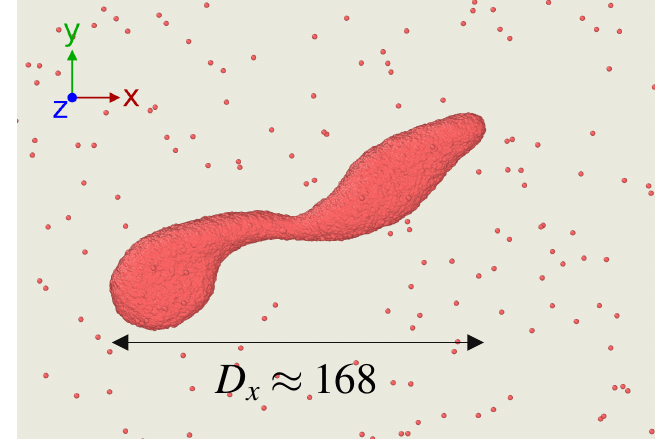} }}%
    \subfloat[\centering ]{{\includegraphics[width=0.33\columnwidth,trim=4.5cm 12cm 6cm 8.5cm,clip]{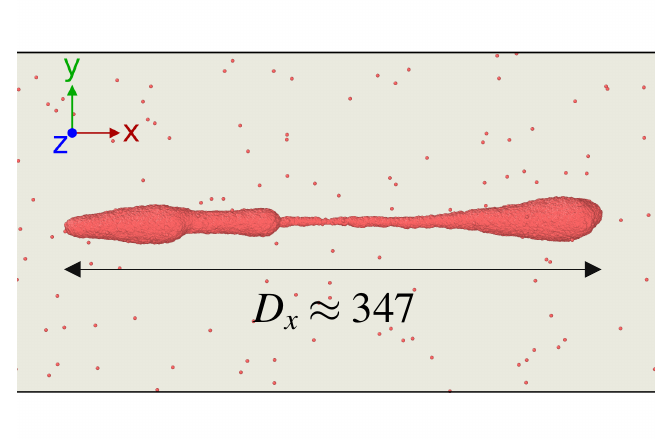} }}%
    %[trim={left bottom right top},clip]
\caption{Top-view snapshots of the three oscillation phases, 
generated using OVITO software.\cite{Stukowski2010}
All droplets have particle number 
$N=20\times10^4$, equilibrium contact angle $\theta=90^{\circ}$ 
(droplet--substrate affinity $\varepsilon_{\rm ws}=1.8$), 
initial contact length along the $x$ direction $D_{x}\approx48$, 
attractive strength $A=-40$, and repulsive strength $B=25$;
see Section~\ref{sec:methods} for details.  
The substrate vibration frequency is fixed at $\omega_{\rm sub}=0.015\pi$, 
with vibration amplitudes: 
(a) $A_{\rm sub}=450$ for phase I (multimedia available online); 
(b) $A_{\rm sub}=675$ for phase II (multimedia available online); 
(c) $A_{\rm sub}=900$ for phase III (multimedia available online). 
Instantaneous contact lengths $D_x$ are shown in the figures. 
}
\label{fig:movies}
\end{figure}

In previous research,\cite{Ng2025} we investigated the oscillations
of water droplets of different sizes on a horizontally vibrating
substrate of varying wettability, across a range of vibration
frequencies and amplitudes, and established the simulation setup that 
allows us to conduct further investigations in this area.
In the case of water droplets, we identified three distinct oscillation
phases. Phase I (Figure~\ref{fig:movies}(a), multimedia available online) is a stable regime in which the droplet maintains a 
generally spherical contact surface and remains synchronized with the 
substrate vibration over an extended period. 
Phase II (Figure~\ref{fig:movies}(b), multimedia available online) is a rotation-dominated regime in which the droplet 
begins to rotate during oscillation, leading to a highly asymmetric
contact surface and eventual droplet breakup. 
Phase III (Figure~\ref{fig:movies}(c), multimedia available online) is a shear-dominated regime in which the droplet undergoes almost 
immediate breakup upon substrate vibration.
Each phase occurs at different amplitudes and frequencies, 
but phase III, which manifests by the
rapid breakup of the droplet due to the oscillations,
was more commonly observed in the case of hydrophilic substrates. 
By harvesting the advantages of the chosen simulation method, we were 
able to characterize the oscillatory behavior in detail
and unveil key properties, such as the critical role of
internal shear stress, which goes beyond the capabilities
of experimental methods. While the influence of the capillary number (Ca) 
emerged from the analysis due to the combined viscous and surface tension effects, where $\mathrm{Ca}=\frac{\mu v}{\sigma}$ for a fluid with velocity $v$, viscosity $\mu$, and surface tension $\sigma$, further data (e.g., simulations
of different liquids) were required to clarify its role in 
describing the different oscillation phases, the
transitions between them, and the various instabilities arising from substrate vibrations, along with their growth rates.
In this direction then, the study of different liquids
becomes indispensable for completing the picture of 
droplet oscillation phenomena in pure liquid droplets on
horizontally (non-axisymmetric) vibrating, flat, solid substrates. 
In addition, the identification of further criteria to characterize 
transitions between the observed droplet phases, such as the emergence
of phase II  oscillations, is highly desired.

We have taken appropriate steps to fill this gap by
carrying out extensive simulations using the MDPD method, 
and we report the results and conclusions here.
Our investigations considered different liquids and a suitable
range of parameters for substrate vibration, 
droplet size, and substrate wettability. Based on detailed
analysis of our results, we have identified the capillary number (Ca) as a suitable 
parameter for discussing the observed phases and their transitions.
In addition, the rotational instability in phase II oscillations has
been identified through analysis of the angular momentum and vorticity, 
which provide information on the growth rate of the developed instability
that leads to the breakup of the droplet.
Moreover, in view of the lack of an explicit potential function 
for the interactions between particles in the MDPD method,
since the force is directly encoded in the equations of motion, 
we have shown that the number of contacts is a reliable measure
analogous to a free-energy-related quantity for 
characterizing the state of the system, and its key components
in the droplet--substrate system,  namely the particle--particle (pp)
and particle--substrate (ps) interactions.
Finally, analysis of various other properties, such as variations in droplet contact length during oscillations and the velocity field, fills out the
picture of this fundamental phenomenon.

In the following section, we provide details of the model and methods used in our research and determine fundamental properties
of the droplet--substrate system, such as the contact angle and the
fundamental frequency. We then present the results
and relevant discussion in Section~\ref{results}. 
Finally, we draw our main conclusions from this work
in Section~\ref{conclusions} and discuss next steps in the area of droplet oscillation phenomena.

\section{Model and Methods}
\label{sec:methods}
The MDPD method has evolved from dissipative particle
dynamics (DPD).\cite{hoogerbrugge1992,Lavagnini2021}
In contrast to DPD, however, it includes both attractive and repulsive (many-body)
interactions. Based on a particular expression for the free
energy,\cite{Vanya2018} it allows the modeling of free
surfaces despite the short range of the interactions.
Hence, it is well suited for the simulation of liquid--vapor
interfaces,\cite{pagonabarraga2000,warren2003}
such as sessile droplets surrounded by vapor. 
In the standard MDPD method, the evolution of particles in
time is described by the Langevin equation (Eq.~\ref{eq1}),
typically producing microscopic states in the canonical ensemble.
This reads for each particle, $i$:
\begin{eqnarray}
m\frac{d\bm{v}_i}{dt} = \sum_{j\neq i} \left(\bm{F}_{ij}^C + \bm{F}_{ij}^R + \bm{F}_{ij}^D\right).
\label{eq1}
\end{eqnarray}
Here, particle $i$ interacts with other particles $j$ in its vicinity 
within a cutoff via a conservative force $\bm{F}^C_{ij}$, and
also experiences a random force $\bm{F}^R_{ij}$ and
a dissipative force $\bm{F}^D_{ij}$.
In contrast to standard Langevin molecular
dynamics, the dissipative force depends on the relative velocity between particles $i$ and $j$.
The integration of Eq.~\ref{eq1} for each particle (coarse-grained bead)
is realized by means of a modified velocity-Verlet
algorithm\cite{groot1997} with a time step, 
$\Delta t = 0.005$ (MDPD units), which is smaller than the typical
MDPD time step of $0.01$ due to the acceleration of the droplet particles
caused by the substrate vibrations. 
Throughout our study, simulations were carried out by using the LAMMPS software.\cite{LAMMPS}
The conservative force is expressed as 
\begin{eqnarray}
\bm{F}^C_{ij} =  Aw_{c}(r_{ij})\bm{e}_{ij} + 
B \left(\bar{\rho}_i + \bar{\rho}_j \right) w_{d}(r_{ij})\bm{e}_{ij},
\label{eq2}
\end{eqnarray}
where $A<0$ determines the strength of the attractive interactions
and $B>0$ that of the repulsive ones.
The repulsive force depends on the local density, 
incorporating many-body effects and satisfying the 
``no-go'' theorem for the force to be conservative.\cite{warren2013}
The distance between particles is denoted by $r_{ij}$,
while $\bm{e}_{ij}$ denotes the unit vector 
from particle \textit{i} to particle \textit{j}. 
The weight functions $w_{c}(r_{ij})$ and $w_{d}(r_{ij})$ are defined as
\begin{eqnarray}
w_{c}(r_{ij}) = 
\begin{cases}
&1 - \frac{r_{ij}}{r_{c}}, \ \ r_{ij} \leq r_{c} \\
& 0,  \  \ r_{ij} > r_{c},
\end{cases} 
\label{eq3}
\end{eqnarray}
\begin{eqnarray}
w_{d}(r_{ij}) = 
\begin{cases}
&1 - \frac{r_{ij}}{r_{d}}, \ \ r_{ij} \leq r_{d} \\
& 0,  \  \ r_{ij} > r_{d}.
\end{cases} 
\label{eq3.2}
\end{eqnarray}
Here, the cutoff distance for the attractive interactions is $r_c=1$,
while $r_d=0.75r_c$ for the repulsive interactions.\cite{Vanya2018} 
The many-body contributions in the repulsive
part of the conservative force are expressed through
the local densities, $\bar{\rho}_i$ and $\bar{\rho}_j$,
which are obtained by the following expression:
\begin{eqnarray}
\bar{\rho}_i = \sum_{j\neq i} \frac{15}{2\pi r_d^3} \left( 1 - \frac{r_{ij}}{r_d}\right)^2.
\label{eq4}
\end{eqnarray}
Thermal fluctuations are implemented in the Langevin thermostat
through the random and dissipative forces, as described in the equation of motion (Eq.~\ref{eq1}),
with the temperature set to unity (MDPD units).
The random and dissipative forces are described by
the following relations:
\begin{eqnarray}
\bm{F}^{D}_{ij} = -\gamma_D w_{D}(r_{ij}) (\bm{e}_{ij} \cdot  \bm{v}_{ij})\bm{e}_{ij} ,
\label{eq5}
\end{eqnarray}
\begin{eqnarray}
\bm{F}^R_{ij} = \sigma_R w_{R}(r_{ij}) \xi_{ij} \Delta t^{-1/2}\bm{e}_{ij},
\label{eq26}
\end{eqnarray}
where $\gamma_D$ is the dissipative strength, $\sigma_R$ 
is the strength of the random force, 
$\bm{v}_{ij}$ is the relative velocity between particles,
and $\xi_{ij}$ is a random 
variable from a Gaussian distribution with zero mean and unit variance. 
The fluctuation--dissipation theorem establishes the
relation between $\gamma_D$ and $\sigma_R$,\cite{espanol1995}
that is
\begin{eqnarray}
\gamma_{D} = \frac{\sigma_{R} ^2}{2 k_B T}, 
\label{eq7}
\end{eqnarray}
while the weight functions for the dissipative and random forces are
\begin{eqnarray}
w_{D}(r_{ij}) = \left[w_{R}(r_{ij})\right]^2 = \left( 1 - \frac{r_{ij}}{r_c}\right)^2.
\label{eq8}
\end{eqnarray}

To simulate a range of different liquids, we vary the 
value of the attractive-force strength, $A$, namely $A=-30,-40,-60$, and $-80$, \cite{zhou2019,Carnevale2024} 
while keeping the 
dissipative coefficient constant at $\gamma_D=4.5$.\cite{Ghoufi2011, arienti2011}
The conversions from reduced to real units for the particle properties 
based on this model for $A=-40$ (water)
are reported in Table~\ref{tab:units}.
\begin{table}[b]
\caption{\label{tab:units} Relation between MDPD units and real units. The scaling 
is performed by matching the surface tension $\gamma$ and density $\rho$ of water to values measured from MDPD
simulations using $A=-40$ and $B=25$. The coarse-graining level is defined so that
one MDPD particle represents three water molecules. For the MDPD examples of other liquids, one may refer to Ref.~\citen{arienti2011}.}
\begin{ruledtabular}
\begin{tabular}{lll}
Parameter  & MDPD value  & Real value \vspace{.1cm} \\ 
 \hline 
 % \vspace{.3cm}
Particle &  1    & 3 H$_2$O  \\ 
$r_c$    &  1     &  8.17 \AA          \\ 
$\rho$   &  6.10  &  997 kg/m$^3$       \\
$\gamma$ &  7.30  &  72 mN/m     \\
%\rho=6.05,\gamma=7.62
\end{tabular}
\end{ruledtabular}
\end{table}
In MDPD, varying $A$ affects both the surface tension, 
$\sigma$, of the liquid--air surface, which is determined according to 
Ref.~\citen{arienti2011}, and the viscosity, $\mu$, of the liquid. 
The exact values of both properties for different $A$ are reported in 
Table~\ref{tab:A_mu_oh}. The combined effect of $\sigma$ 
and $\mu$, for 
a given characteristic length scale $\ell$ (e.g., suspended droplet radius $R$), can be 
expressed through the Ohnesorge number (Oh):
\begin{eqnarray}
\textrm{Oh} = \frac{\mu}{\sqrt{\rho\sigma \ell}}.
\label{eqOh}
\end{eqnarray}
As the magnitude of attractive strength $|A|$ becomes higher, Oh shows a clear upward trend, 
and therefore a greater role of the viscous
forces compared to surface tension. 
Therefore, in the following we will often imply that an 
increased droplet viscosity is equivalent to larger values of $|A|$.
\begin{table}[b]
\caption{\label{tab:A_mu_oh} Density ($\rho$), surface tension ($\sigma$), dynamic viscosity ($\mu$), suspended droplet radii ($R$), and Ohnesorge number ($\rm{Oh}$) for various attractive strengths $A$, with the repulsive strength fixed at $B=25$ and the particle number fixed at $N=20\times10^4$, in the MDPD model.}
\begin{ruledtabular}
\begin{tabular}{llllll}
$A$  & $\rho$  & $\sigma$  &  $\mu$  & $R$ & $\rm{Oh}$\vspace{.1cm} \\ 
 \hline 
 % \vspace{.3cm}
-30 &  5.2629  &  4.5343   & 3.4009   & 20.8564 &  0.1524\\ 
-40 &  6.0902  &  8.1041   & 5.1982   & 19.8657 &  0.1660\\ 
-60 &  7.6109  &  19.0050  & 10.7600  & 18.4432 & 0.2083 \\
-80 &  9.1011  &  36.2551  & 33.9000  & 17.376  & 0.4477 \\
\end{tabular}
\end{ruledtabular}
\end{table}
In the context of our work and the system under investigation, 
it is natural to consider the capillary number (Ca), which accounts for both viscous and surface 
tension effects in the \textit{moving} (oscillating) droplet. 
The exact range of Ca explored in our study depends on the 
equilibrium contact angle of the droplet, which is denoted
by $\theta$. For the droplets examined here, the size
is above the threshold at which the contact 
angle is independent of it.\cite{theodorakis2015modelling}
For the simulation setup presented here, where 
horizontal substrate oscillations occur along the $x$ direction, the capillary number
$\rm Ca$ is defined as:
\begin{eqnarray}
\textrm{Ca} = \frac{\mu |\bar{u}_{x}|}{\sigma},
\label{eqCa}
\end{eqnarray}
where the characteristic velocity $|\bar{u}_{x}|$ 
is the mean particle velocity in the $x$ direction, measured along the central line $y=0$ at the contact surface located at $z=10$.

\begin{figure}[htb!]
\includegraphics[width=0.6\columnwidth,trim=5cm 5.5cm 4.8cm 8cm,clip]{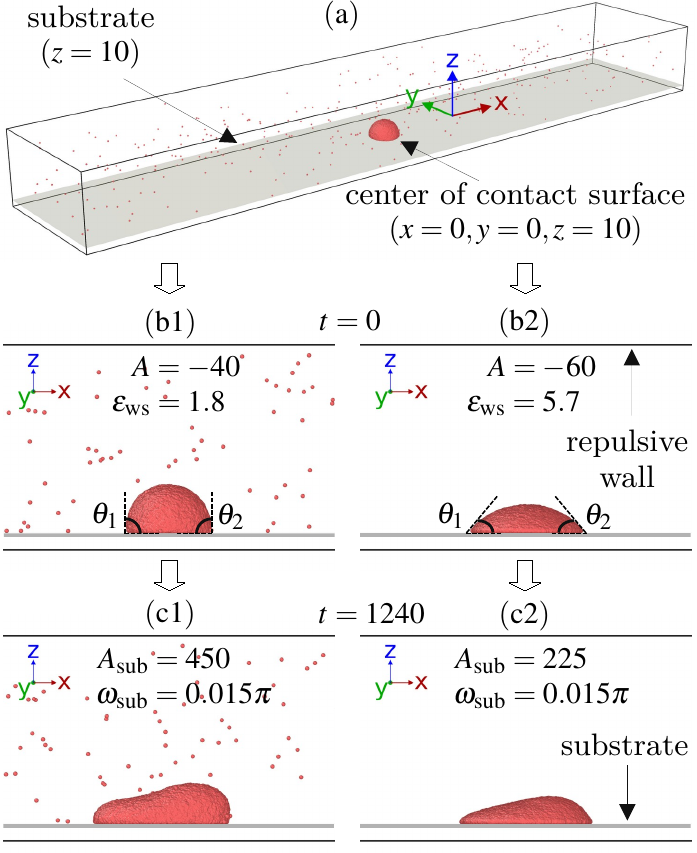}%[trim={left bottom right top},clip]
\caption{Snapshots of the simulation setup generated using OVITO software. \cite{Stukowski2010} 
(a) Perspective view of a static droplet on a substrate
(see Eq.~\ref{eq:LJpotential93}). 
The substrate is set at the plane $z=10$, and the droplet is initially 
positioned at the center ($x=0, y=0$).
(b1,b2) Front views of the simulation. 
The equilibrium contact angle of the droplet is adjusted by tuning the
attractive strength $A$ and the affinity $\varepsilon_{\rm{ws}}$
(see Table~\ref{tab:epsilon}). 
In this example, (b1) uses $A=-40, \varepsilon_{\rm{ws}}=1.8$;
(b2) uses $A=-60, \varepsilon_{\rm{ws}}=5.7$.
These parameter sets result in different static contact angles
$\theta_{1}$ and $\theta_{2}$ at equilibrium state
(for each static case, of course, $\theta_{1}=\theta_{2}=\theta$). 
(c1,c2) A time-dependent sinusoidal vibration 
(see Eq.~\ref{eq:substrate_velocity}) is applied to the substrate, 
with amplitude $A_{\rm{sub}}$ and frequency $\omega_{\rm{sub}}$, 
inducing droplet oscillation. Both snapshots are taken at $t=1240$.
Here, both cases use the same frequency $\omega_{\rm sub}=0.015\pi$; 
(c1) has an amplitude $A_{\rm sub}=450$, 
while (c2) has $A_{\rm sub}=225$.}
\label{fig:setup}
\end{figure}

To set up the simulations for sessile equilibrium and 
oscillating droplets, freely suspended equilibrium droplets
were initially generated, which, in turn, were placed on the solid substrate and allowed to relax (Figure~\ref{fig:setup}(b1) and~\ref{fig:setup}(b2)). After reaching the equilibrium state, a droplet with a lower magnitude of attractive interaction $|A|$ coexists with a denser vapor of free particles.
Typical droplet oscillatory motions corresponding to 
phase I, II, and III are 
illustrated in Figure~\ref{fig:movies}(a) (multimedia available online), 
Figure~\ref{fig:movies}(b) (multimedia available online), and 
Figure~\ref{fig:movies}(c) (multimedia available online), respectively.
Our main investigations were 
carried out with droplets of approximately $N=20\times10^4$ MDPD
particles. We also performed simulations with droplets containing up to 
$N=10^6$ particles to examine the dependence of the
fundamental frequency on droplet size in droplet--substrate systems
and to assess possible finite-scale effects. The typical radius
of a droplet with $N=20\times10^4$ particles is about $24$ (MDPD units).
The droplet was placed at the center of the substrate, with the
coordinates of the center-of-mass of the droplet coinciding
with coordinates $x=0$ and $y=0$, while the substrate plane was located at $z=10$. To ensure that the observed three 
oscillation phases are solely due to the momentum transfer from the
non-axisymmetric oscillation of the substrate to the 
droplet particles, and to avoid any factors related to the 
roughness of the substrate, such as pinning, the substrate was modeled as a smooth, unstructured, flat solid
wall. This was taken into account implicitly via a Lennard-Jones (LJ) 9--3 potential,\cite{Theodorakis_MCVOF_2021}
implemented in LAMMPS as follows:
\begin{equation}\label{eq:LJpotential93}
U^{\rm 9-3}(z') = \varepsilon_{\rm ws} \left[  \frac{2}{15}\left(\frac{\sigma_{\rm  ws}}{z'}
\right)^{9} - \left(\frac{\sigma_{\rm ws}}{z'}  \right)^{3}    \right],
\end{equation}
where the 9--3 form comes from the integration of the well-known
12--6 LJ potential.\cite{Israelachvili2011,Forte2014} The interaction cutoff of the substrate
with the particles was $r_{\rm LJ}=2.93$. The particles experienced a force 
$F_{\rm ws}=-\frac{\partial U^{\rm 9-3}(z')}{\partial z'}$, 
where $z'$ is the normal distance between each particle and
the substrate. The effective diameter of the Lennard-Jones potential was set to $\sigma_{\rm ws}=1.17$.
The parameter $\varepsilon_{\rm ws}$ was used to tune the
affinity of the droplet to the substrate, which in turn
controlled the contact angle of the droplet (substrate wettability), 
since the solid--vapor interfacial tension is expected to be 
negligible.\cite{Tretyakov2014}
Values of $\varepsilon_{\rm ws}$ ranged from $0.4$ to $9.0$, 
yielding equilibrium contact angles of sessile droplets $\theta=50^\circ$,
$90^\circ$, $110^\circ$, and $130^\circ$ (see Table~\ref{tab:epsilon} for each specific case).

\begin{table}[b]
\caption{\label{tab:epsilon} Values of droplet--substrate affinity $\varepsilon_{\rm{ws}}$ for different attractive strengths $A$, with a fixed repulsive strength $B=25$, and the corresponding contact angles $\theta$ in the MDPD model.}
\begin{ruledtabular}
\begin{tabular}{lllll}
$A$  & $\varepsilon_{\rm{ws}} (\theta=50^{\circ})$  & $\varepsilon_{\rm{ws}} (\theta=90^{\circ})$  &  $\varepsilon_{\rm{ws}} (\theta=110^{\circ})$ &  $\varepsilon_{\rm{ws}} (\theta=130^{\circ})$\\ 
 \hline 
 % \vspace{.3cm}
$-30$ &  $1.9$  & $1.2$  &  $0.7$  &  $-$   \\ 
$-40$ &  $3.0$  & $1.8$  &  $1.0$  &  $0.4$ \\ 
$-60$ &  $5.7$  & $3.0$  &  $2.0$  &  $-$   \\
$-80$ &  $9.0$  & $5.0$  &  $3.0$  &  $-$   \\
\end{tabular}
\end{ruledtabular}
\end{table}

This range of contact angles covers both 
hydrophilic and hydrophobic substrate cases, which
are relevant experimentally \cite{Gilewicz_Master_Thesis} 
and suitable for MDPD simulations, since, for example,
significant contact line distortion or precursor film formation for small contact angles is avoided.
The contact angle was determined by fitting the liquid--vapor interface profile with a polynomial. In particular,
we obtained the surface curvature of the $y=0$ cross-section
of the droplet by locating the coordinates of the outermost MDPD particles at the cross-section and fitting the particles with the polynomial:
\begin{equation}\label{eq:polynomial}
x = p(z) = p_{1}z^{n}+p_{2}z^{n-1} + \ldots + p_{n}z+p_{n+1},
\end{equation}
where the coefficients $p_{n}$ were obtained via
least-squares fitting. In line with our previous study,\cite{Ng2025}
we found that values $3\leq n\leq5$ yield the 
best fit for the droplet surface profile 
without distortion near the contact line. 
Following this procedure, 
we measured the tangent angles from the
sequential tangents to the polynomial function (Eq.~\ref{eq:polynomial}), starting from the fitting point 
nearest to the contact line, $z\gtrsim10$, and moving to higher $z$. The equilibrium contact angle $\theta$ is determined until 
the change in tangent angle is below a set threshold value, $\Delta\theta$. 
Here, we set $\Delta\theta = 1.5^{\circ}$, which is less than the 
standard deviations of the randomness of the angles, $2^{\circ}$, as noted in our previous study.\cite{Ng2025}

While the contact angle in the case of the equilibrium sessile
droplet is expected to be the same on average axisymmetrically,
namely around the droplet, advancing and receding contact angles differ in the case of oscillating droplets, depending on
the direction of motion. 
Hence, we denote $\theta_1$ as the
left contact angle (negative $x$ direction) and $\theta_2$ as the right contact angle
(positive $x$ direction) of the droplet, 
respectively.
The standard deviations of the randomness in the equilibrium
contact angles, $\sigma_{\theta_{1}}$ and $\sigma_{\theta_{2}}$, 
due to statistical fluctuations and fitting errors, were of the order of $2^{\circ}$, 
consistent with our previous study.\cite{Ng2025}

Apart from monitoring the contact angles and comparing 
them with the equilibrium contact angle, to characterize our droplet--substrate system we also
determined
the fundamental frequency (first breathing mode) of the equilibrium 
sessile droplets. This was done by measuring the oscillation of the droplet's center-of-mass 
in the $z$ direction (axisymmetric system) for each case of $N$ and $\theta$. 
The attractive strength was set to $A=-40$, while the contact angle was
tuned by adjusting the strength of the 9--3 LJ potential, $\varepsilon_{\mathrm{ws}}$, as discussed above.
For a freely suspended droplet, the fundamental frequency mode ($n_{f}=2$) 
is expected to scale as $f_2\sim \sqrt{\ell^{-3}}$ with prefactors depending
on definitions and simplifications made, and with $\ell$ naturally chosen 
as the droplet radius, $R$. \cite{Rayleigh1879} 
While certain studies have attempted to investigate
the Rayleigh spectrum of drops on solid supports,\cite{Strani_Sabetta_1984,Strani1988,Bisch1982, Bostwick_Steen_2013_part1,Bostwick_Steen_2013_part2,Bostwick2009,Fayzrakhmanova2009,Lyubimov2006} the determination 
of the fundamental (lowest) frequency mode of a sessile droplet 
has not been determined. 
In this regard, challenges relate to the lack of spherical symmetry that 
would facilitate descriptions expressing deviations from this ideal
spherical symmetry based on harmonic functions, such as those
used in Rayleigh theory,\cite{Rayleigh1879} and from the difficulty in determining
boundary conditions, especially at the three-phase contact line. The latter
would also require taking into account various effects, such as sliding and
pinning/depinning at the contact line, which would result in mixed 
conditions. Among others, viscous dissipation remains challenging for
theoretical treatments, which generally consider inviscid fluids. For these reasons,
we have chosen to obtain estimates of the fundamental 
frequency directly from computer simulations, aiming to provide 
trends in the lowest frequency mode as a 
function of droplet size and equilibrium contact angle. 

\begin{figure}[htb!]
\includegraphics[width=0.5\columnwidth,trim=3.5cm 3.5cm 4cm 3cm,clip]{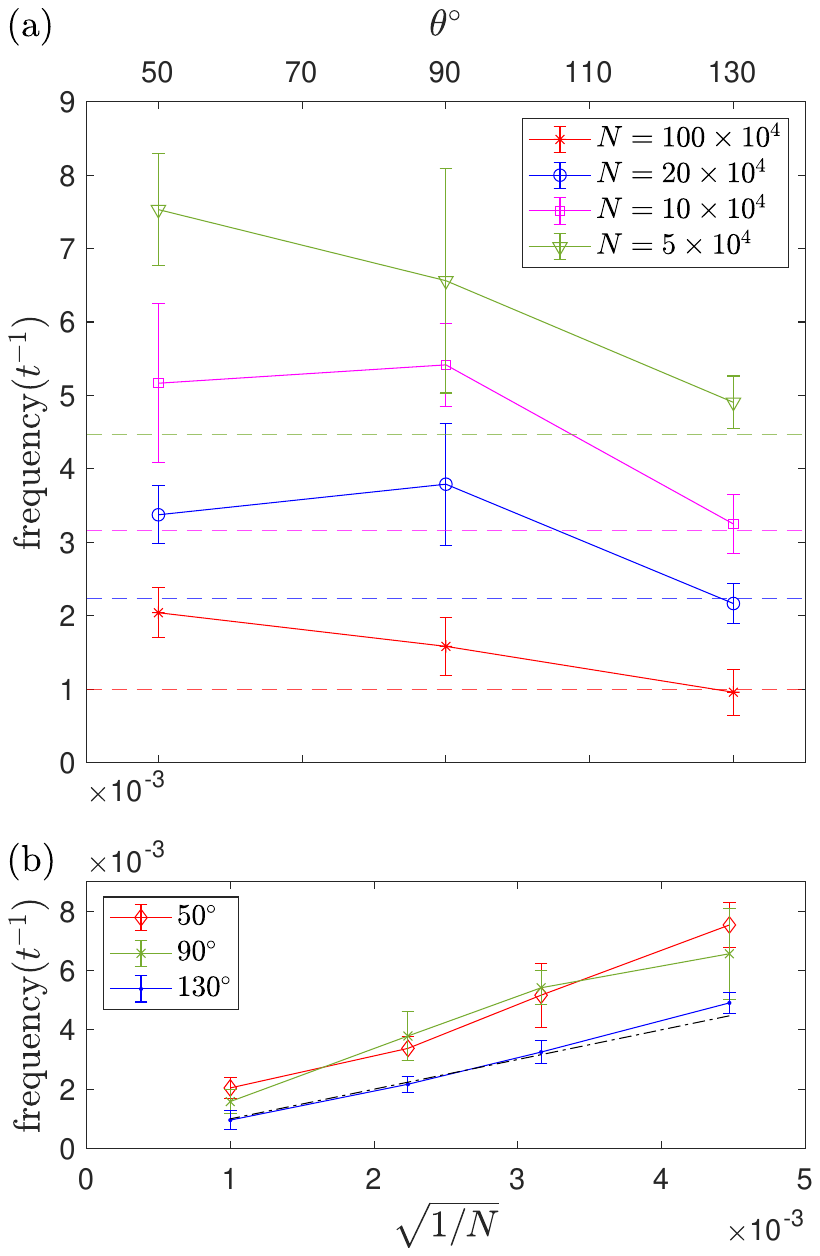}%[trim={left bottom right top},clip]
\caption{Frequency of the fundamental mode ($n_{f} = 2$) for static droplets with $A=-40$ and $B=25$, for particle numbers $N=5\times10^4, 10\times10^4,20\times10^4$ and $100\times10^4$, each at contact angles $\theta=50^{\circ},90^{\circ}$ and $130^{\circ}$. All results are averaged over four independent runs, with error bars representing the standard deviation of the measured frequencies. (a) Horizontal dashed lines indicate the expected scaling $f_{2}\propto\sqrt{N^{-1}}$, showing good agreement for highly spherical droplets ($\theta=130^{\circ}$) when $N\ge10\times10^4$. (b) Black dashed line indicates the expected scaling $f_{2}\propto\sqrt{N^{-1}}$.}
\label{fig:freqz_dcomz}
\end{figure}

The results in Figure~\ref{fig:freqz_dcomz} indicate that the fundamental
frequency of the droplets, describing deviations from an ideal 
spherical shape, decreases for less wettable substrates
and for larger droplets.
Interestingly, we find that droplets with an equilibrium
contact angle of $130^\circ$ and sufficient size ($N\ge10\times10^4$)
follow the expected scaling for freely suspended
droplets, namely $f_2\propto\sqrt{N^{-1}}$ for $\ell^{3}\propto N$, 
suggesting a negligible effect of the presence of the substrate
on the fundamental frequency of the droplet. In contrast, 
larger deviations are observed for droplets with much 
smaller equilibrium contact angles, namely
$\theta=50^\circ$.
Although the droplet's fundamental oscillation is a quantity of 
theoretical interest for describing the oscillations,
the analysis of our results, which will be discussed in subsequent
sections, has not indicated a direct connection between the
fundamental frequency and the observed simulation outcomes 
regarding the behavior of the oscillations, such as
the presence of phases I, II, and III, or other properties such
as velocity fields.
This might be attributed to the large difference in the fundamental
frequency and the frequencies (and amplitudes) in the 
vibrating experiments.
For this reason, henceforth,
the discussion of our results will not be put in the perspective of
the droplet's fundamental frequency.

Turning now our attention to the substrate vibrations, these
occur in the horizontal ($x$) direction with an
amplitude and frequency which remain constant during the
simulation. The instantaneous substrate velocity is given by
\begin{equation}\label{eq:substrate_velocity}
u_{\rm{sub}}(t) = -A_{\mathrm{sub}}\omega_{\mathrm{sub}}\mathrm{sin}(\omega_{\mathrm{sub}}t),
\end{equation}
where $A_{\mathrm{sub}}$ is the vibration amplitude and
$\omega_{\mathrm{sub}}=2\pi f_{\mathrm{sub}}$ is the angular frequency,
with $f_{\mathrm{sub}}$ being the vibration frequency.
For the results presented in this work, amplitudes and frequencies 
were chosen in the ranges $225\leq A_{\rm sub}\leq1500$ and
$0.0075\pi\leq\omega_{\rm sub}\leq0.3\pi$, respectively. 
These values span the range of relevant 
experimental parameters for non-axisymmetric vibration 
experiments.\cite{Gilewicz_Master_Thesis} 

To realize the simulation experiments, the freely suspended, 
equilibrated droplet is first placed on the substrate.
Following a relaxation toward the system's equilibrium state, 
the substrate vibrations are activated.
The total simulation time of each simulation varies depending on the observed 
behavior, ensuring that the time scale is 
sufficient to arrive at conclusions. 
Typically, this simulation time ranges from
$5000 \leq t \leq 10000$ (in MDPD units, natural for the model).
Average properties for static cases were obtained from 
ensembles of statistically independent samples,
while in the case of oscillating droplets, properties were
continuously monitored with an interval of $\Delta t=2.5$.

\section{Results}
\label{results}

\begin{figure}[htb!]
\includegraphics[width=0.99\columnwidth,trim=3cm 0.2cm 3cm 0.8cm,clip]{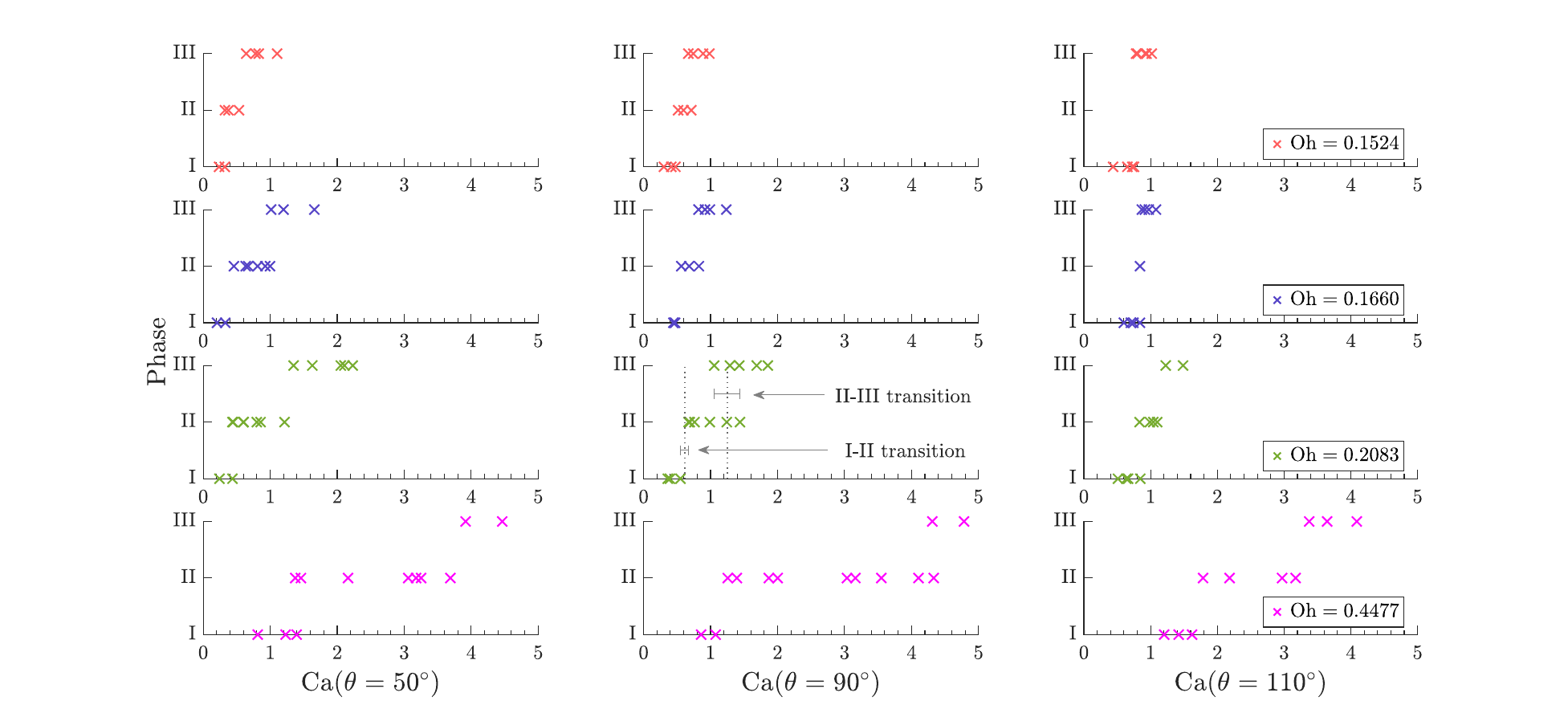}%[trim={left bottom right top},clip]
\caption{Oscillation phases (I, II, III) plotted against the capillary number (Ca) for droplets with various attractive strengths ($\rm{Oh}=0.1524,0.1660,0.2083,$ and $0.4477$) and equilibrium contact angles ($\theta=50^\circ$, $90^\circ$, and $110^\circ$), as indicated. 
The number of particles is fixed at $N=20\times10^4$ for all cases. Each data point corresponds to a distinct set of substrate vibration amplitude $A_{\rm sub}$ and frequency $\omega_{\rm sub}$.
}
\label{fig:theta_all_ca_phase}
\end{figure}

\begin{figure}[htb!]
\includegraphics[width=0.8\columnwidth,trim=2.1cm 7.3cm 2.6cm 7.2cm,clip]{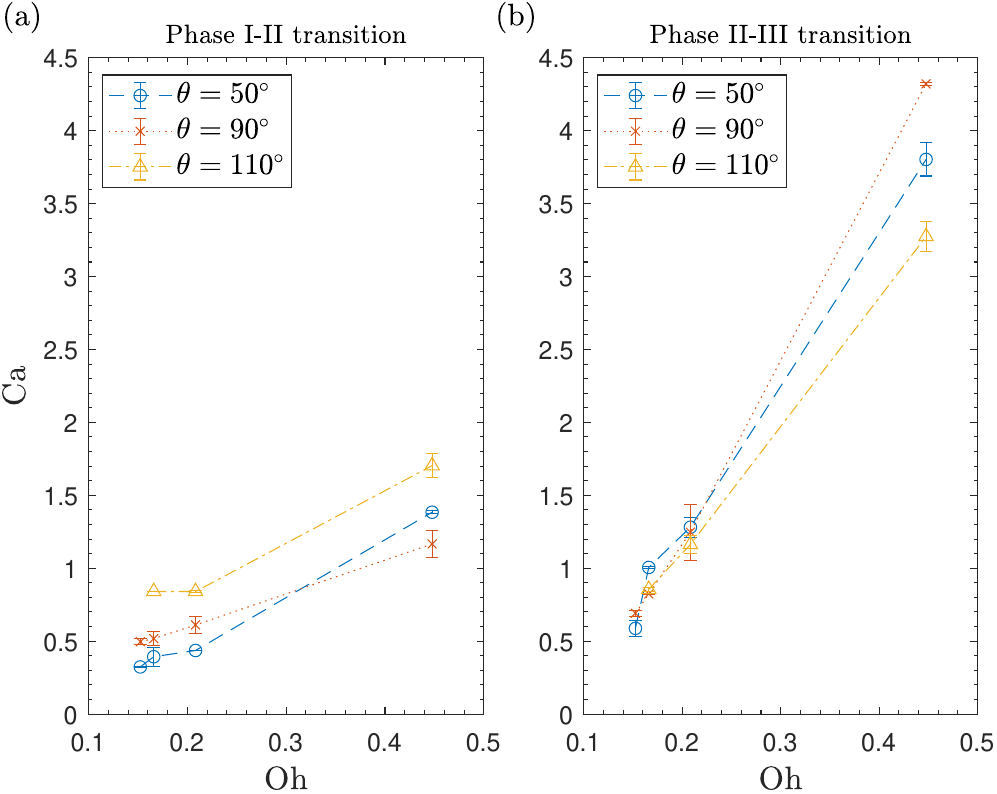}%[trim={left bottom right top},clip]
\caption{Capillary number (Ca) for phase transitions I--II (a)
and II--III (b) plotted against the 
Ohnesorge number $(\rm{Oh})$ for various equilibrium contact angles
$\theta$ (see Figure \ref{fig:theta_all_ca_phase}).
Error bars indicate the overlapping range of Ca for each transition.
}
\label{fig:phase_transition}
\end{figure}

\subsection{Oscillation Phases and the Capillary number}

We examined the occurrence of phases I, II, and III as a function of the capillary number
(Ca) to discuss and categorize the droplet 
oscillation motion in a universal way. The oscillating droplet exhibits different
phases depending on both the vibration amplitude, $A_{\rm sub}$, and 
frequency, $\omega_{\rm sub}$.
Turning our attention to a specific angle, 
for example, $\theta=50^\circ$, we find that an increasing $\rm{Oh}$,
which in practice promotes viscous forces over surface tension,
extends the range of Ca in which phase II occurs to higher
values (Figure~\ref{fig:theta_all_ca_phase}).
In general, the transition from phase I to phase II
occurs over a small range of Ca for small $\rm{Oh}$,
and the same is true for the transition from phase II to
III. For less wettable substrates, 
such as $\theta=110^\circ$, we find that this range of Ca is even smaller than
the other wettability cases, and, for $\rm{Oh}=0.1524$, we observe the complete absence of phase II. To provide a more concise view
of these scenarios, we summarize our findings
in Figure~\ref{fig:phase_transition}, where 
the critical Ca values for phase I--II and II--III
transitions are plotted against $\rm{Oh}$.
We find that the critical Ca for the phase I--II transition
clearly increases for more wettable substrates. 
Moreover, regardless of the equilibrium contact angle $\theta$, droplets with higher viscosity (e.g., $\rm{Oh}>0.2083$) 
show a significant increase in the onset of instability (critical Ca value), 
which otherwise remains nearly constant for $\rm{Oh} \leq 0.2083$.
For the phase II--III transition, an increased critical Ca is found with increasing $\rm{Oh}$ across the entire range. 
This critical Ca also shows a relatively significant increase for $\rm{Oh}>0.2083$. 
In this transition, we observe a relatively smaller difference between the various substrates for $\rm{Oh}\leq0.2083$; 
that is, for all $\theta$ cases, the critical Ca values are similar for given $\theta$ and $\rm{Oh}$.
This may indicate that, in this lower $\rm Oh$ regime, the transitions are dominated by the presence of 
the substrate and its vibrations, while the interactions between the 
particles within the droplet play a lesser role in determining the 
oscillatory behavior of the droplets.
As we show later, velocity profile analysis reveals that 
the onset of phase III is driven primarily by shearing of 
the droplet rather than true oscillatory motion, as we 
have also observed in the
case of water droplets,\cite{Ng2025} which may explain the observations.

\begin{figure}[htb!]
{\includegraphics[width=0.5\columnwidth,trim=3.8cm 5.7cm 3.3cm 5.5cm,clip]{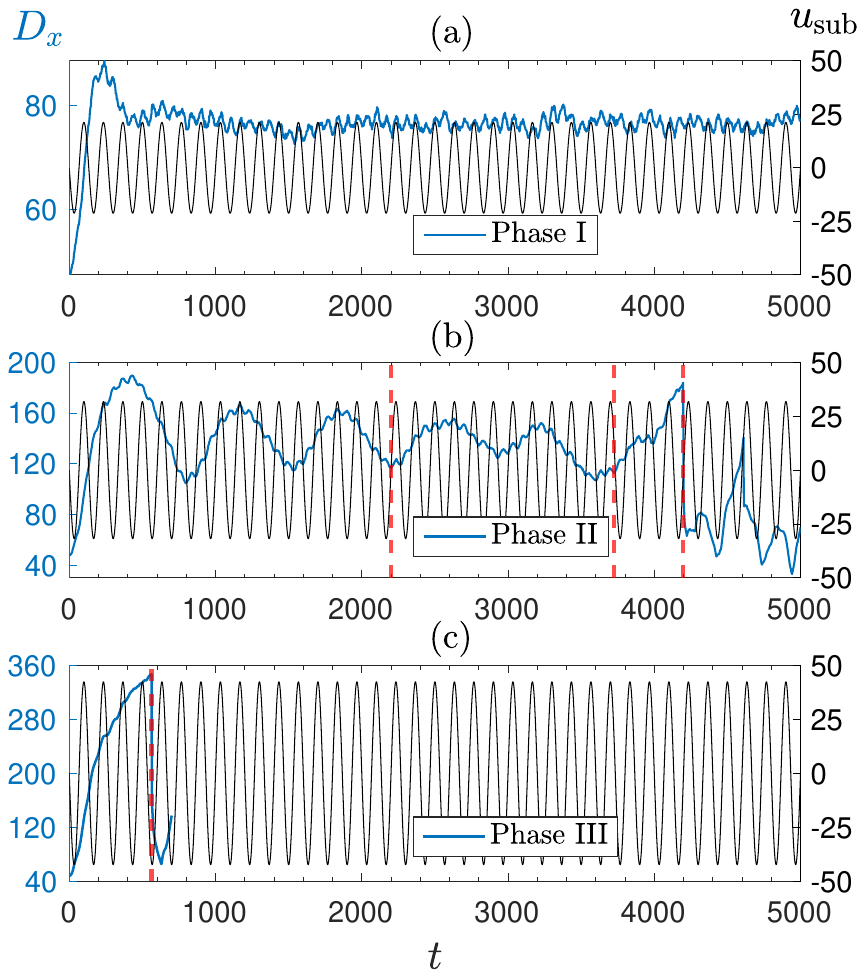} }%
    %[trim={left bottom right top},clip]
\caption{Contact length, $D_{x}$, of the oscillating droplet along the $x$ direction. 
(a) $D_{x}$ for the example of phases I oscillation 
(Figures~\ref{fig:movies}(a), \ref{fig:phase1_scalar}, and \ref{fig:phase1_amz}). 
(b) $D_{x}$ for phase II oscillation 
(Figures~\ref{fig:movies}(b), \ref{fig:phase2_scalar}, and \ref{fig:phase2_amz}).
(c) $D_{x}$ for phase III oscillation 
(Figures~\ref{fig:movies}(c), \ref{fig:phase3_scalar}, and \ref{fig:phase3_amz}). 
Red dashed lines indicate the same time points as in the
corresponding examples.
}
\label{fig:combine_dia}
\end{figure}

\subsection{Rotation dynamics}\label{sec:rotation}
The appearance of phase II is signaled by a 
breaking of symmetry in the oscillatory motion of the droplet,
leading to its rotation and eventually to its breakup 
(see Figure~\ref{fig:movies}(b)
for an example of phase II oscillation, multimedia available online). 
One can therefore consider the $z$-component of the angular momentum,
$L_{z}=r_{x}v_{y}-r_{y}v_{x}$ (for unit particle mass, $m=1$), 
and the vorticity, $\omega_{z}=\partial_{x}v_{y}-\partial_{y}v_{x}$,
as parameters for characterizing the emergence of phase II oscillation 
and its growth. In this section, we present several properties monitored in all three phases: 
the contact length of the droplet, namely $D_x$; 
the average velocity along the $x$ direction
at the contact surface, $\overline{u}_{\rm cs}$; the 
center-of-mass position of the droplet in the $y$ direction, 
$y_{\rm com}$; the average $z$-component of the vorticity at the contact surface, 
$\omega_{z}$; and the average $z$-component of the angular momentum at the contact surface,
$L_{z}$.

\begin{figure}[htb!]
\captionsetup[subfigure]{labelformat=empty}
    \subfloat[\centering ]
    {{\includegraphics[width=0.48\columnwidth,trim=3.7cm 6.0cm 3.3cm 5.5cm,clip]{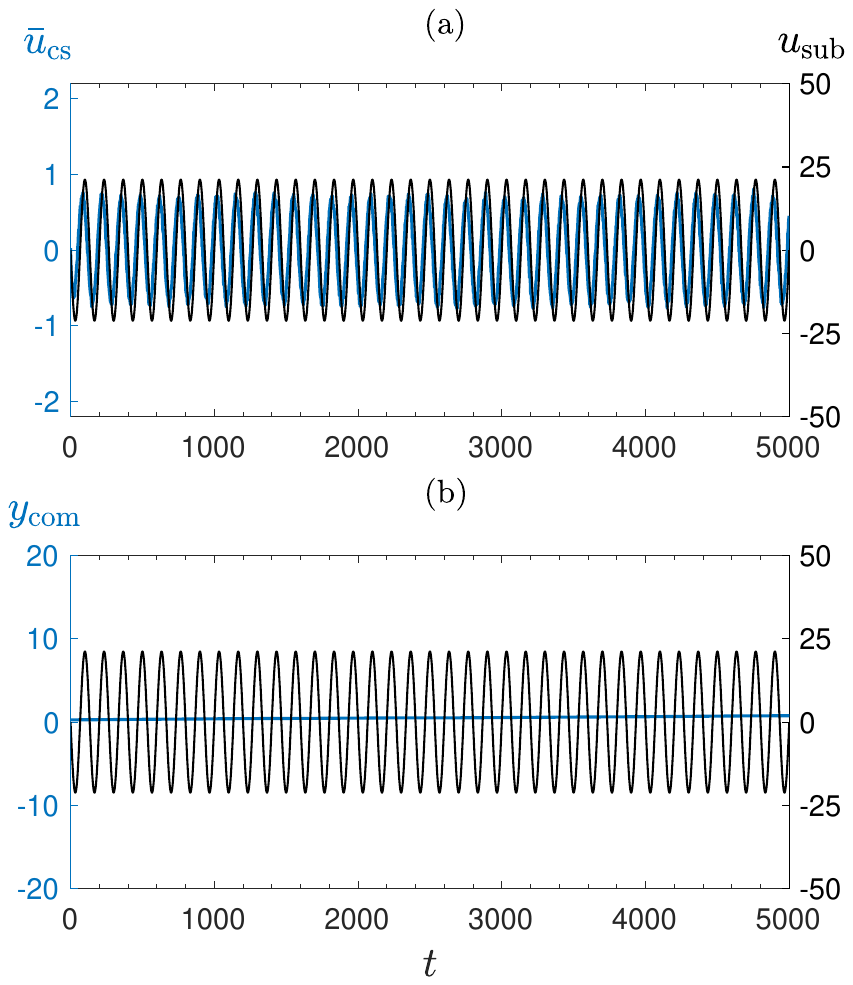} }}%
    %\quad
    \subfloat[\centering ]{{\includegraphics[width=0.48\columnwidth,trim=3.7cm 6.0cm 3.3cm 5.5cm,clip]{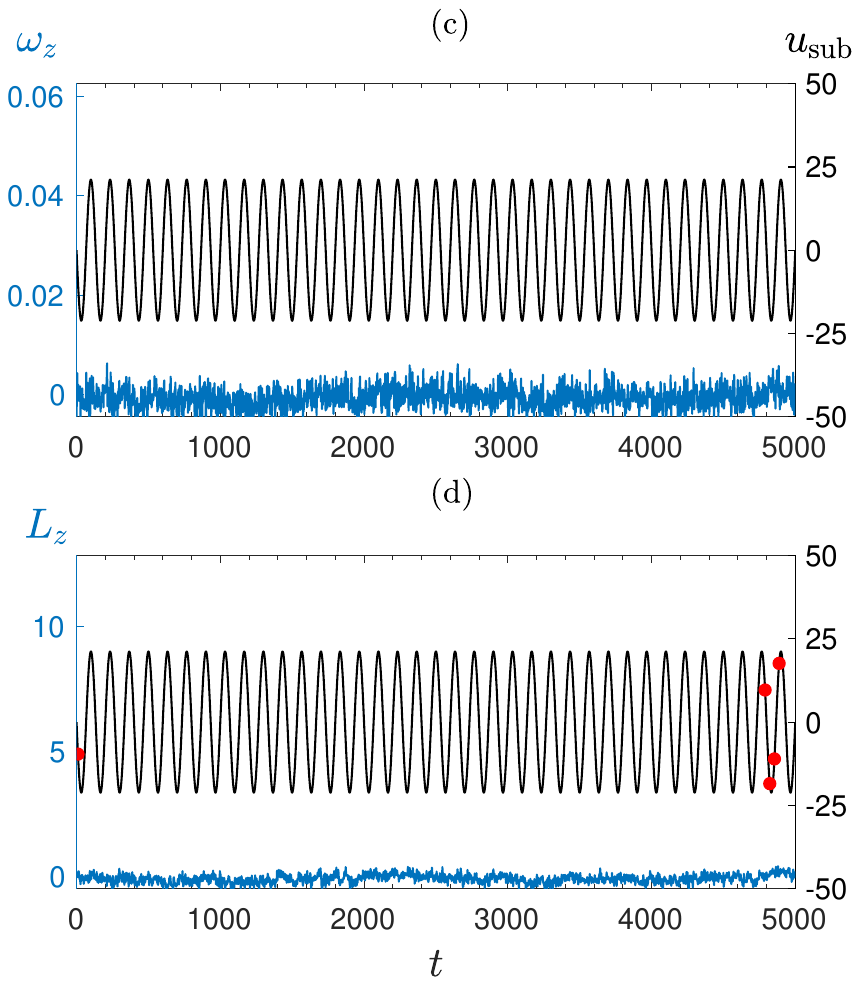} }}%
    %[trim={left bottom right top},clip]
\caption{Oscillation phase I of a droplet at $\rm{Oh}=0.1660$,
with equilibrium contact angle $\theta=90^{\circ}$ and particle number $N=20\times10^4$. 
Here, substrate amplitude $A_{\rm{sub}}=450$, 
substrate frequency $\omega_{\rm{sub}}=0.015\pi$. 
(a) Average velocity of the droplet along the $x$ direction at the contact surface, $\overline{u}_{\rm cs}$; 
(b) center-of-mass of the droplet in the $y$ direction, $y_{\rm com}$; 
(c) average $z$-component of the vorticity of the droplet at the contact surface, $\omega_z$;
(d) average $z$-component of the angular momentum at the contact surface, $L_z$. The red dot markers correspond
to the times of the snapshots in Figure~\ref{fig:phase1_amz}. 
Droplet quantities in blue; substrate velocity $u_{\rm sub}$ in black.
}
\label{fig:phase1_scalar}
\end{figure}

\begin{figure}[htb!]
    \subfloat[\centering ]{{\includegraphics[height=4.3cm,trim=6cm 8.5cm 8cm 8.2cm,clip]{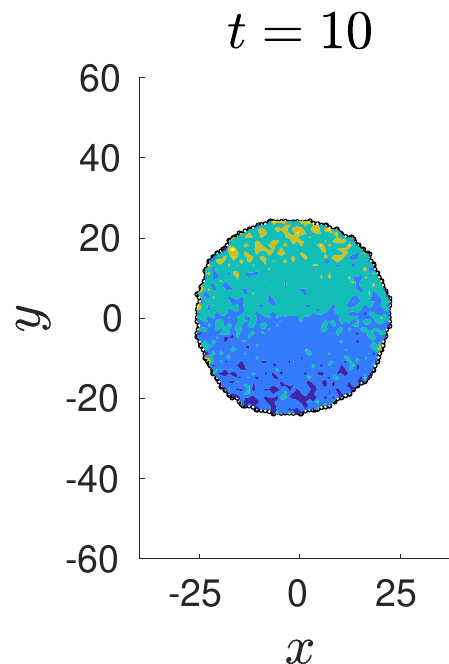} }}%[trim={left bottom right top},clip]
    \quad
    \subfloat[\centering ]{{\includegraphics[height=4.3cm,trim=7cm 8.5cm 6.5cm 8.2cm,clip]{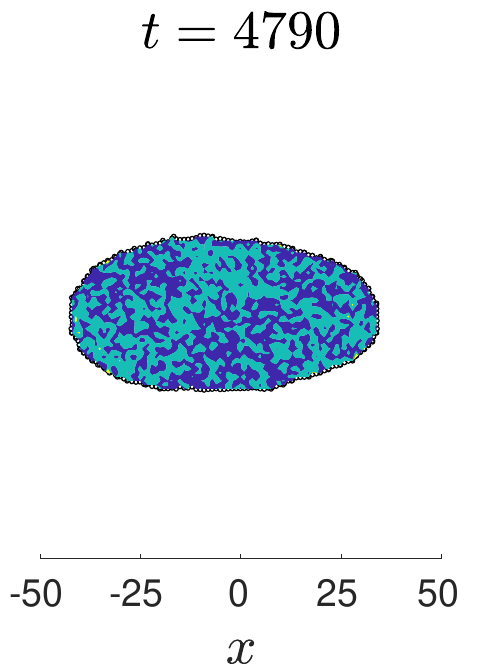} }}%
    \quad
    \subfloat[\centering ]{{\includegraphics[height=4.3cm,trim=7.5cm 8.5cm 7cm 8.2cm,clip]{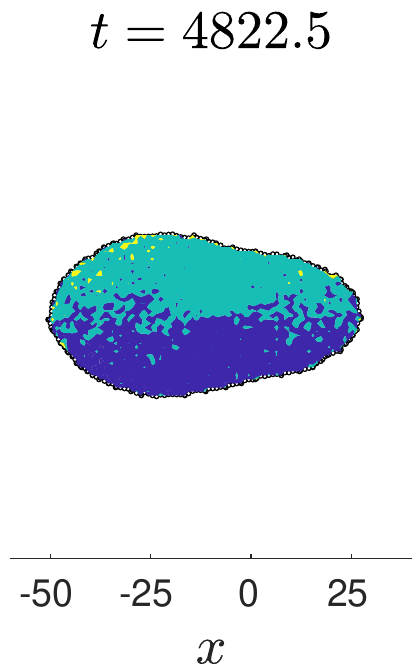} }}%
    \quad
    \subfloat[\centering ]{{\includegraphics[height=4.3cm,trim=7.5cm 8.5cm 7cm 8.2cm,clip]{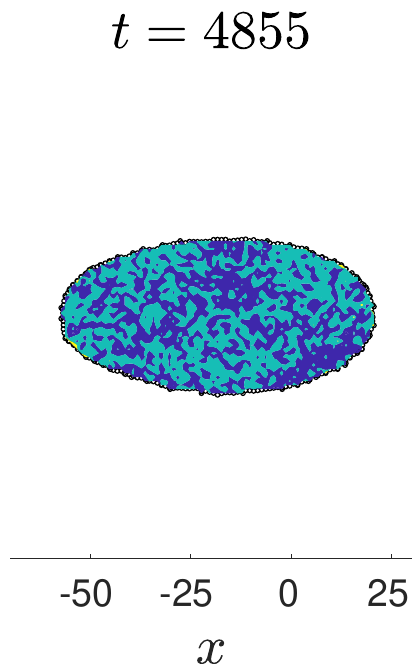} }}%
    \quad
    \subfloat[\centering ]{{\includegraphics[height=4.3cm,trim=7.5cm 8.5cm 7cm 8.2cm,clip]{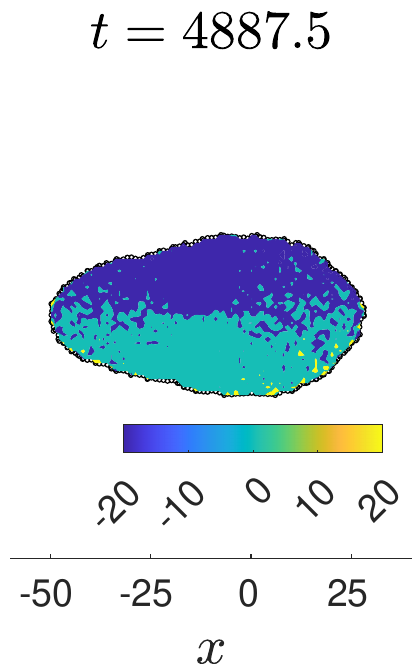} }}%\\
\caption{Profile of the angular momentum, $L_{z}$, of the droplet at the contact
surface during phase I oscillation (see also Figure~\ref{fig:movies}(a), multimedia available online). The times $t$ correspond to the red 
dot markers in Figure~\ref{fig:phase1_scalar}(d). 
The color scale reflects the magnitude of $L_{z}$.
}
\label{fig:phase1_amz}
\end{figure}

In the case of phase I oscillations, 
after the initial extension of the droplet's contact length, $D_x$, 
the droplet re-contracts to a contact length that is stable over time, 
with slight contraction–extension synchronized with 
the horizontal vibrations of the substrate (Figure~\ref{fig:combine_dia}(a)). 
The droplet oscillations also closely follow the substrate vibrations throughout the entire simulation, 
as can clearly be observed in the behavior of 
$\overline{u}_{\rm cs}$ (Figure~\ref{fig:phase1_scalar}(a)). Moreover, the droplet's motion occurs along
the $x$ direction on an almost perfectly straight line throughout the
simulation, with the center-of-mass position remaining approximately constant, namely $y_{\rm com} \approx 0$ (Figure~\ref{fig:phase1_scalar}(b)). 
In this case, the angular momentum and vorticity
fluctuate around zero (Figures~\ref{fig:phase1_scalar}(c) and~\ref{fig:phase1_scalar}(d)), indicating that an overall rotational motion
is negligible in phase I oscillation, and the momentum 
transfer from the vibrating substrate to the droplet takes place solely in the vibration direction. 
The deformations of the droplet caused by the substrate vibrations 
almost preserve the spherical symmetry of the droplet, as shown in Figure~\ref{fig:phase1_amz}. 
The time points chosen in this phase I example,
$t=4790$, $4822.5$, $4855$, and $4887.5$, 
correspond to one complete cycle of substrate vibration near the end
of the simulation ($t=0$ corresponds to the start
of the horizontal substrate vibrations). In Figures~\ref{fig:phase1_amz}(b)--\ref{fig:phase1_amz}(e), the angular momentum distribution 
at each time remains symmetric across the regions $y<0$ and $y>0$, 
and the sequential patterns exhibit 
almost perfect periodicity.
In phase I oscillation, no sustained symmetry breaking develops as in phase II discussed in the following. 
In phase III oscillation (discussed in the following after phase II), the symmetry is also preserved 
up to breakup, but the droplet dynamics are qualitatively different from phase I.

\begin{figure}[htb!]
\captionsetup[subfigure]{labelformat=empty}
    \subfloat[\centering ]{{\includegraphics[width=0.48\columnwidth,trim=3.7cm 5.6cm 3.3cm 5.5cm,clip]{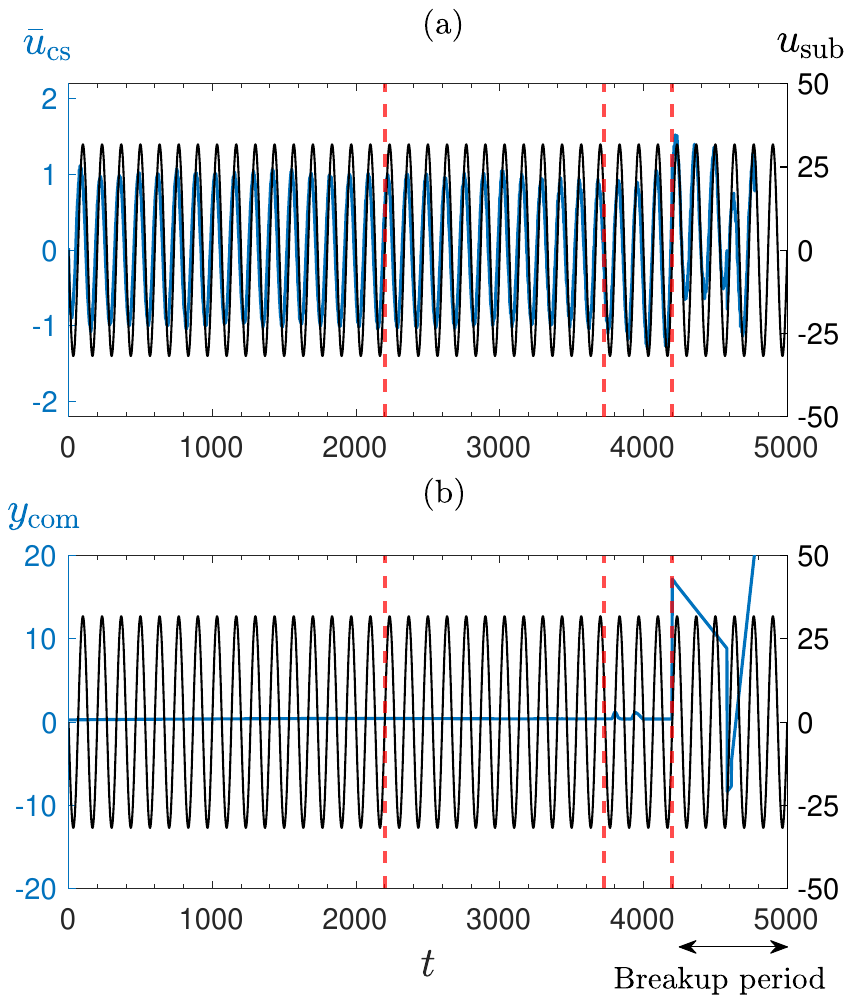} }}%
    %\quad
    \subfloat[\centering ]{{\includegraphics[width=0.48\columnwidth,trim=3.7cm 5.6cm 3.3cm 5.5cm,clip]{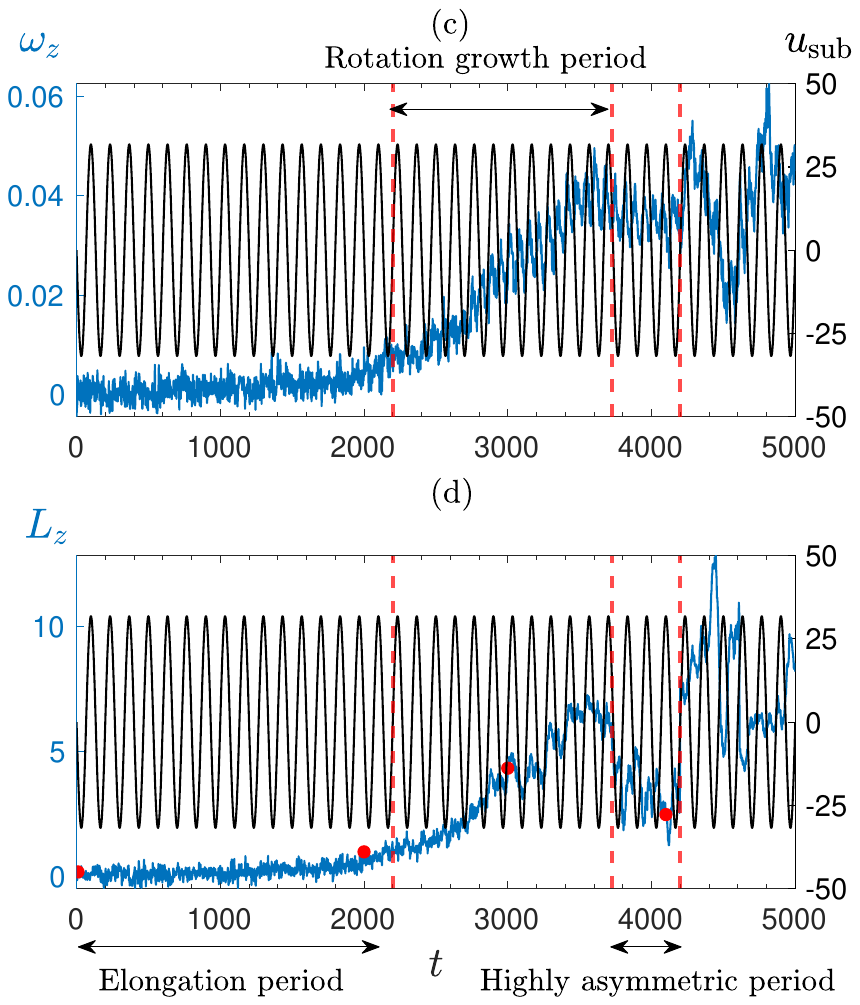} }}%
    %[trim={left bottom right top},clip]
\caption{Oscillation phase II of a droplet at $\rm{Oh}=0.1660$.
The droplet has an equilibrium contact angle $\theta=90^{\circ}$ with
$N=20\times10^4$ particles. 
Here, substrate amplitude $A_{\rm{sub}}=675$ and frequency $\omega_{\rm{sub}}=0.015\pi$. 
(a) Average velocity of the droplet along the $x$ direction at the contact surface, $\overline{u}_{\rm cs}$; (b) center-of-mass of the droplet in the $y$ direction, $y_{\rm com}$; 
(c) average $z$-component of the vorticity of the droplet at the contact surface, $\omega_z$;
(d) average $z$-component of the angular momentum at the contact surface, $L_z$. 
The red dot markers correspond
to the times of the snapshots in Figure~\ref{fig:phase2_amz}. 
Red dashed lines indicate the times at which: 
the droplet becomes elongated ($t<2200$); asymmetry develops due to 
rotation ($2200<t<3700$); and droplet breakup occurs ($t=4200$).
Droplet quantities in blue; substrate velocity $u_{\rm sub}$ in black.
}
\label{fig:phase2_scalar}
\end{figure}

\begin{figure}[htb!]
    \subfloat[\centering ]{{\includegraphics[height=4.3cm,trim=3.4cm 8.2cm 11.2cm 8.5cm,clip]{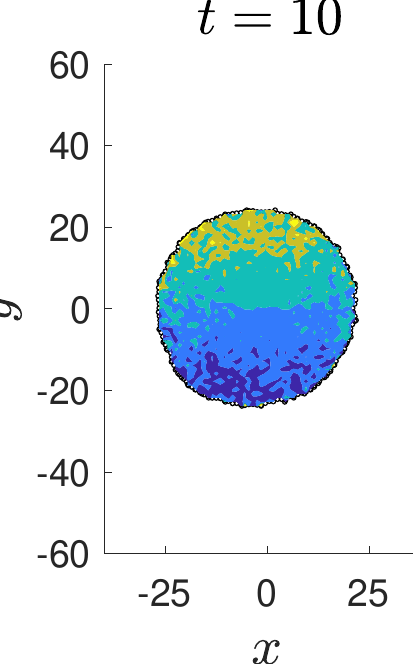} }}%[trim={left bottom right top},clip]
    \quad
    \subfloat[\centering ]{{\includegraphics[height=4.3cm,trim=6.3cm 8.2cm 5.5cm 8.5cm,clip]{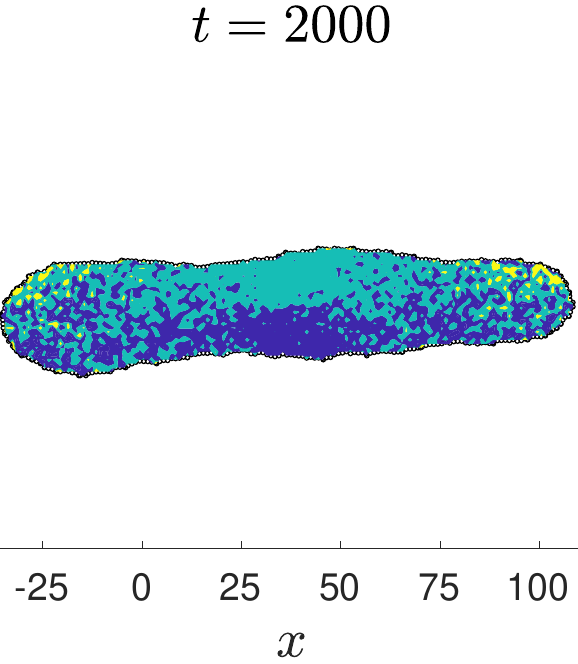} }}%
    \quad
    \subfloat[\centering ]{{\includegraphics[height=4.3cm,trim=6.5cm 8.2cm 5.5cm 8.5cm,clip]{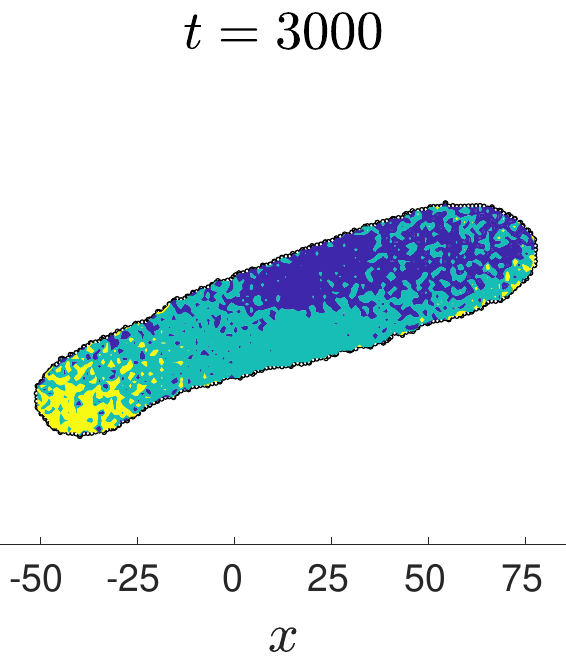} }}%
    \quad
    \subfloat[\centering ]{{\includegraphics[height=4.3cm,trim=5.5cm 8.2cm 4.5cm 8.5cm,clip]{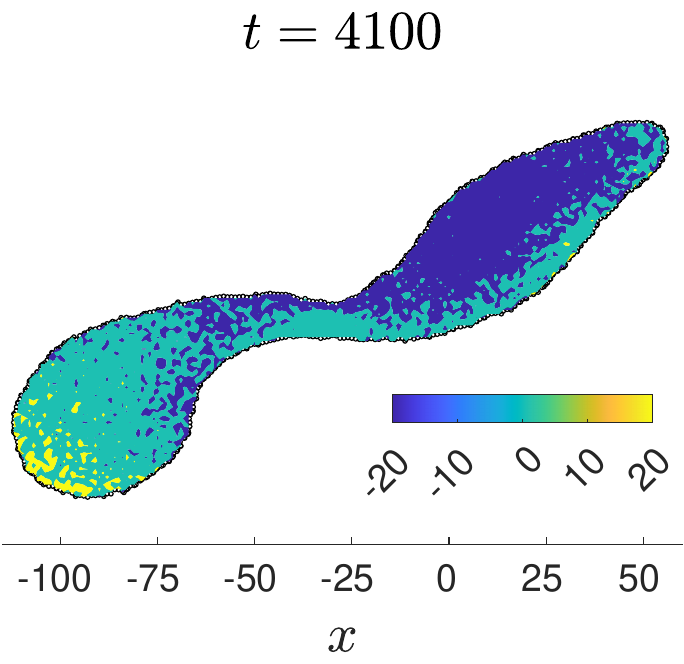} }}%\\
\caption{Profile of the angular momentum $L_{z}$ of the droplet
at the contact surface during phase II oscillation, prior to breakup (see also Figure~\ref{fig:movies}(b), multimedia available online). 
The times $t$ correspond to the red dot markers in Figure~\ref{fig:phase2_scalar}(d).
The color code reflects the magnitude of $L_z$.
}
\label{fig:phase2_amz}
\end{figure}

A typical behavior of phase II oscillations is showcased
in Figure~\ref{fig:phase2_scalar}. 
During the interval $0<t<2200$, the droplet
oscillates in a regularly repeating pattern.
In practice, this means that the average velocity along the $x$ direction
at the contact surface, $\overline{u}_{\rm cs}$, follows the 
pattern of the substrate vibrations (Figure~\ref{fig:phase2_scalar}(a)) 
and the droplet
moves along the $x$ direction of the vibrations
preserving its center-of-mass position in the $y$ direction,
i.e., $y_{\rm com}\approx0$ (see Figure~\ref{fig:phase2_scalar}(b)). 
The droplet is elongated in the direction of vibration 
($x$ direction), maintaining a symmetric shape about the central 
line of the droplet at $y=0$ (Figure~\ref{fig:phase2_amz}(b)).
We refer to this initial stage of phase II oscillation 
as the \textit{elongation period}, as illustrated in the example 
of Figure~\ref{fig:phase2_scalar}. During this period,
both the angular momentum and vorticity of the droplet
remain close to zero, indicating that the rotation motion
of the particles is minimal (Figures~\ref{fig:phase2_scalar}(c) and~\ref{fig:phase2_scalar}(d)). 
In addition, the elongated, oscillating droplet exhibits a metastable breathing mode,
undergoing several 
cycles of contraction and extension with a period longer than 
that of the substrate vibrations (see Figure~\ref{fig:movies}(b), multimedia available online). In the example shown here, the droplet completes three 
contraction--extension cycles during the elongation period 
(see Figure~\ref{fig:combine_dia}(b)). This periodic change in contact length, $D_x$,
is also reflected in the number of particle contacts 
within the droplet, which will be discussed later in 
Section~\ref{sec:contacts}. Over time, these contraction--extension cycles 
gradually increase the uneven mass distribution within the droplet,
eventually creating a sufficiently asymmetric angular-momentum 
distribution. Then, during the interval $2200<t<3700$, 
the angular momentum and vorticity of the droplet begin to grow, marking the 
\textit{rotation growth period} (Figures~\ref{fig:phase2_scalar}(c) and~\ref{fig:phase2_scalar}(d)). In this stage,
a global rotation of the droplet develops due to the growing
asymmetry in the angular-momentum distribution at the 
contact surface, which is more clearly illustrated in 
Figure~\ref{fig:phase2_amz}. Here, the particles near 
the substrate are illustrated at various times
marked by the red dot markers in Figure~\ref{fig:phase2_scalar}(d). 
As seen in Figures~\ref{fig:phase2_amz}(a)--\ref{fig:phase2_amz}(b), the angular momentum is initially
symmetric about $y=0$ 
with opposite magnitudes during the elongation period. 
This symmetry is later broken, as shown in 
Figures~\ref{fig:phase2_amz}(c)--\ref{fig:phase2_amz}(d).
During the interval $3700<t<4200$, the rotating droplet
eventually reaches a highly asymmetric state, 
characterized by local rotations within the droplet,
culminating in a sudden change in both angular momentum
and vorticity at $t=3700$ (in this example, both $L_{z}$ and 
$\omega_{z}$ drop at $t=3700$ 
as shown in Figures~\ref{fig:phase2_scalar}(c) and~\ref{fig:phase2_scalar}(d)). 
The highly asymmetric rotating droplet 
may also experience shifts in its center-of-mass due to the uneven mass 
distribution, as indicated by the two small peaks in $y_{\rm com}$
at approximately $t\approx3800$ and $t\approx3950$ in
Figure~\ref{fig:phase2_scalar}(b). This asymmetry eventually leads to irregular oscillations and droplet breakup
at $t=4200$, which is reflected across all properties shown 
in Figure~\ref{fig:phase2_scalar}. Moreover, we track the
dominant portion of the droplet after the breakup.
We find that the sudden changes in the center-of-mass position,
$y_{\rm com}$, the vorticity, $\omega_z$, and the angular momentum, 
$L_{z}$, at $t=4200$, are due to the uneven mass distribution 
of the smaller rotating child droplet. This scenario is 
reflected in the sharp change of $y_{\rm com}$ with a shift of the center-of-mass ($|\Delta y_{\rm com}|\approx17$).
Finally, we observe that the droplet can rotate either clockwise or counterclockwise
in different simulation runs. 
The duration of each
oscillation stage and the degree of asymmetry also vary depending on the initial conditions,
in this case, the initial velocities of the particles at the moment the
substrate vibration is activated.

\begin{figure}[htb!]
\captionsetup[subfigure]{labelformat=empty}
    \subfloat[\centering ]
    {{\includegraphics[width=0.48\columnwidth,trim=3.7cm 5.6cm 3.3cm 5.5cm,clip]{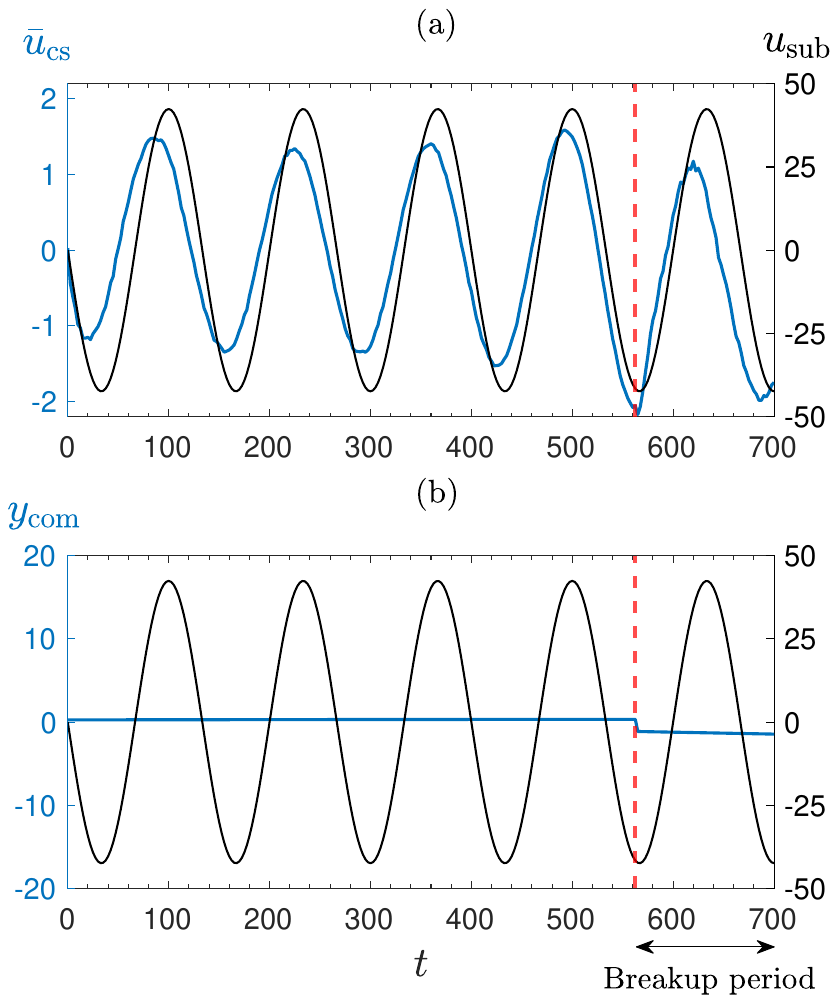} }}%
    %\quad
    \subfloat[\centering ]{{\includegraphics[width=0.48\columnwidth,trim=3.7cm 5.6cm 3.3cm 5.5cm,clip]{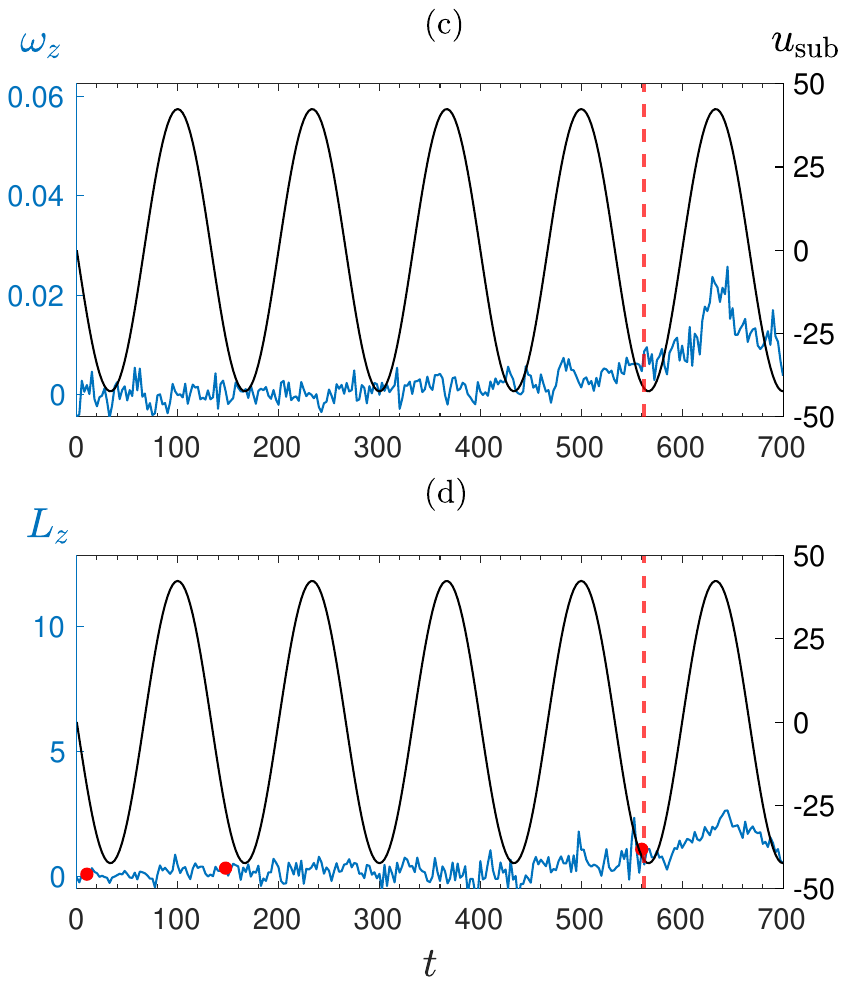} }}%
    %[trim={left bottom right top},clip]
\caption{Oscillation phase III of a droplet at $\rm{Oh}=0.1660$, with equilibrium contact angle $\theta=90^{\circ}$,
particle number $N=20\times10^4$, substrate amplitude $A_{\rm{sub}}=900$, and frequency $\omega_{\rm{sub}}=0.015\pi$. 
(a) Average velocity of the droplet along the $x$ direction at the contact surface, $\overline{u}_{\rm cs}$;  
(b) center-of-mass of the droplet in the $y$ direction, $y_{\rm com}$; 
(c) average vorticity of the droplet at the contact surface, $\omega_{z}$; 
(d) average angular momentum of the droplet at the contact surface, $L_{z}$. Red dashed line indicates the time $t=562$, when droplet breakup occurs, while the red dot markers correspond to the times of the snapshots in Figure~\ref{fig:phase3_amz}.
}
\label{fig:phase3_scalar}
\end{figure}

\begin{figure}[htb!]
    \subfloat[\centering ]{{\includegraphics[height=8.4cm,trim=2.5cm 6.5cm 14.5cm 7.5cm,clip]{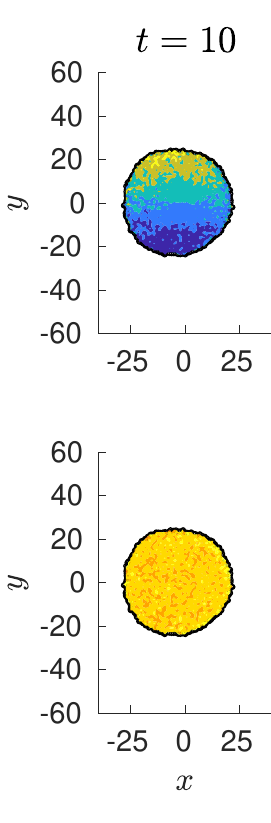} }}%[trim={left bottom right top},clip]
    \subfloat[\centering ]{{\includegraphics[height=8.4cm,trim=7cm 6.5cm 7cm 7.5cm,clip]{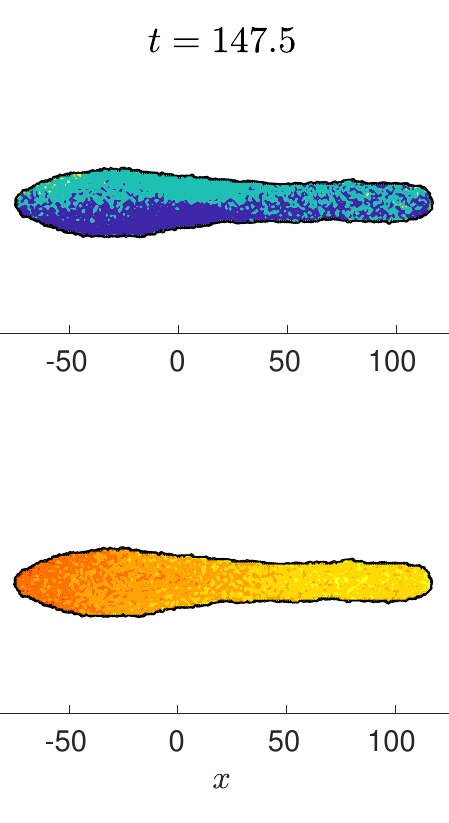} }}%
    \subfloat[\centering ]{{\includegraphics[height=8.4cm,trim=3.5cm 6.5cm 4cm 7.5cm,clip]{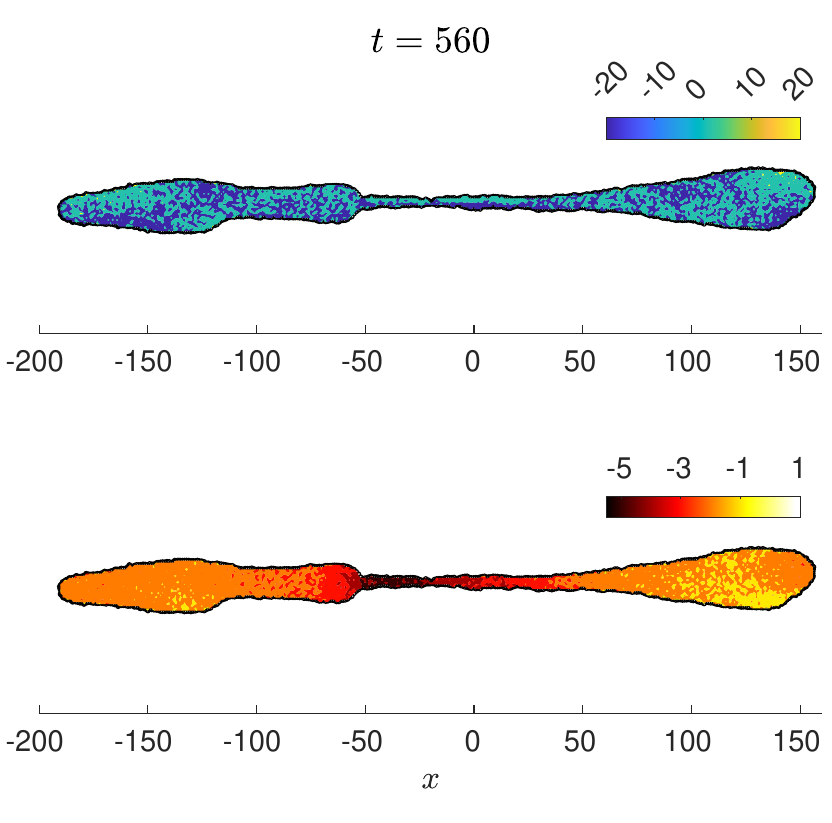} }}%

\caption{Upper panel: profile of the angular momentum $L_{z}$ of the droplet at the contact surface during phase III oscillation 
(see also Figure~\ref{fig:movies}(c), multimedia available online). 
Lower panel: corresponding profile of the local velocity of the droplet along the $x$ direction at the contact surface, $u_{\rm cs}$. 
The times $t$ correspond to the red dot markers
in Figure~\ref{fig:phase3_scalar}(d).
}
\label{fig:phase3_amz}
\end{figure}

The case of phase III oscillations exhibits a markedly different behavior from phases I and II (Figure~\ref{fig:phase3_scalar}) and
represents an extreme out-of-equilibrium scenario.
Here, regular oscillatory motion following
that of the substrate vibrations is largely absent,
as indicated by the phase lag in $\overline{u}_{\rm cs}$ 
with respect to the substrate vibrations (Figure~\ref{fig:phase3_scalar}(a)).  
The droplet is in a highly non-equilibrium state, which
points to a scenario where the droplet is being shaken
rather than oscillating. This is clearly indicated
by the short time required for the droplet breakup to occur 
from the initial extension of the contact length (Figure~\ref{fig:combine_dia}(c)); in the example shown, $t=562$. 
The angular momentum and vorticity remain close to zero (Figures~\ref{fig:phase3_scalar}(c) and~\ref{fig:phase3_scalar}(d)), 
showing only a slight increase just before breakup. 
Similar to phase II oscillations, when tracking the largest 
child droplet after breakup, we observe a sudden shift in the 
center-of-mass position $y_{\rm com}$ at the breakup moment (Figures~\ref{fig:phase3_scalar}(b)). 
However, since no global rotation is involved in the breakup during
phase III oscillation, the shift in the center-of-mass ($|\Delta y_{\rm com}|\approx1.5$) is much 
smaller than that observed in phase II.
Characteristic profiles of the angular momentum at different time points are shown
in Figure~\ref{fig:phase3_amz}, along with the corresponding profiles of local velocity 
along the $x$ direction at the contact surface, $u_{\rm cs}$.
The angular momentum distribution exhibits symmetry across the 
regions $y<0$ and $y>0$, with opposite magnitudes on either side.
In contrast to the breakup in phase II,
this symmetry is preserved even up to the moment just before breakup, 
as shown in Figure~\ref{fig:phase3_amz}(c). 
Meanwhile, the velocity profile reveals a significant difference between the
center and the left $(x\lesssim0)$ and right $(x\gtrsim0)$ ends of the droplet.
Once breakup occurs, the droplet in this example splits at the 
center into two parts of roughly equal size.

\begin{figure}[htb!]
\includegraphics[width=0.8\columnwidth,trim=1.7cm 7.2cm 2.1cm 7.0cm,clip]{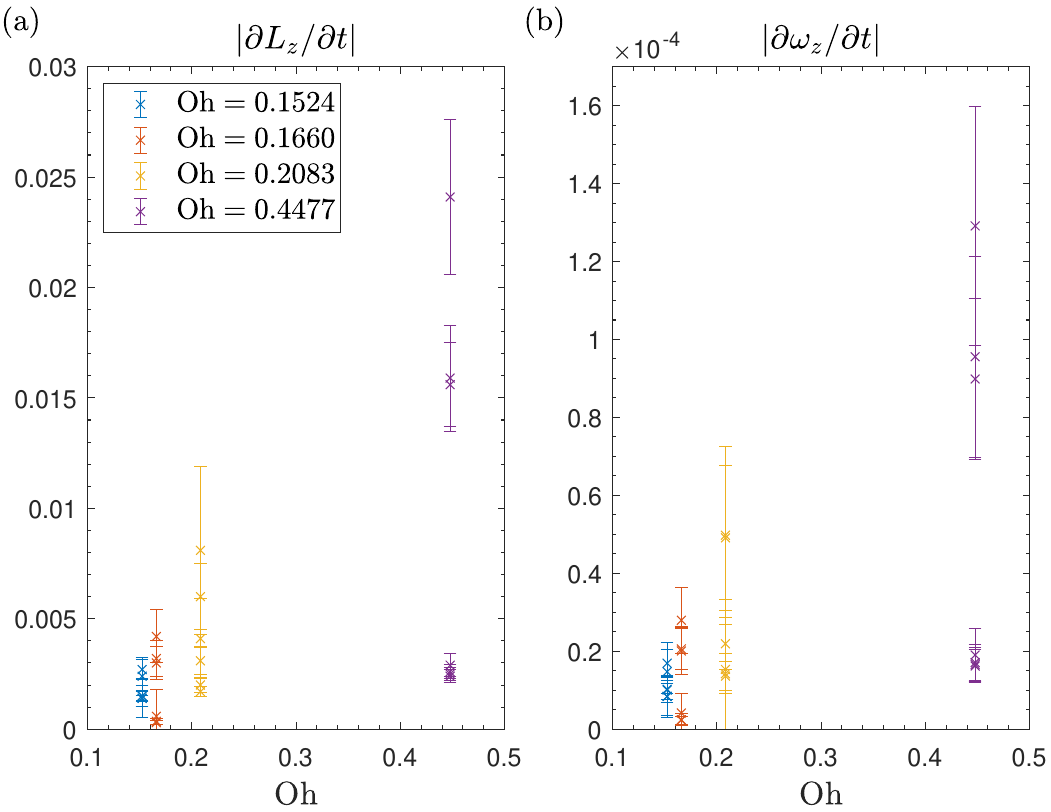}
\caption{Growth rate magnitudes of angular momentum $|\partial L_z/\partial t|$ (a) 
and vorticity $|\partial\omega_z/\partial t|$ (b) 
for droplets in phase II oscillation
at various Ohnesorge numbers ($\rm{Oh}$). 
All droplets have a static contact angle $\theta=90^{\circ}$ and a particle number $N=20\times10^4$.
}
\label{fig:growth_rate}
\end{figure}

\subsection{Growth rate of the rotation}
Having chosen the angular momentum, $L_{z}$, and vorticity, $\omega_{z}$, as
parameters for characterizing the rotational motion
of the droplet that leads to its breakup in phase II,
one can go a step further by calculating the growth
rate of the developing instability via their
time derivatives. A least-squares fitting method is applied to obtain the 
growth rates of both quantities during the rotation growth period in phase
II oscillation.
For the example shown in Figures~\ref{fig:phase2_scalar}(c) and~\ref{fig:phase2_scalar}(d), 
the growth rates are determined over the time interval $2200<t<3700$.
In Figure~\ref{fig:growth_rate}, we plot the magnitudes of
both growth rates, namely $|\partial L_z/\partial t|$
and $|\partial\omega_z/\partial t|$, as functions of the
Ohnesorge number $\rm{Oh}$, where a higher $\rm{Oh}$ corresponds to a more viscous liquid (i.e., stronger interparticle interactions), 
as indicated by the corresponding higher attractive strength magnitude $|A|$, 
listed in Table~\ref{tab:A_mu_oh}.
Here, a droplet with equilibrium contact angle 
$\theta=90^\circ$ and particle number $N=20\times10^4$
is chosen as an example, but 
conclusions are similar for the other cases. 
To determine the range of growth rate magnitudes, we selected sets of 
vibration amplitudes $A_{\rm sub}$ and frequencies 
$\omega_{\rm sub}$ that induce phase II oscillation, 
near the phase I--II and phase II--III transitions.
We find that the maximum growth rate magnitude increases with increasing
values of $\rm{Oh}$. In other words, as viscous forces play
a more pronounced role, 
the ranges of vorticity and angular momentum growth rate magnitudes increase substantially,
since a wider range of vibration frequencies and amplitudes
is favorable for phase II oscillation. 
This finding regarding the two growth rates of rotation aligns 
with the observed changes in the capillary number Ca for phase 
II, shown in Figure~\ref{fig:theta_all_ca_phase},
where increasing $\rm{Oh}$ extends the range of Ca 
for phase II oscillation.
Moreover, the maximum vorticity growth rate magnitude of the droplet 
increases as the viscous forces become more dominant,
due to the enhanced momentum transfer between particles 
arising from the stronger interparticle interactions. 
This observation also aligns with the increased maximum 
values of Ca for phase II with increasing $\rm{Oh}$, 
as shown in Figure~\ref{fig:theta_all_ca_phase}.
Overall, the results indicate that more viscous fluids exhibit a broader
range of rotational growth rates. This is because higher viscosity makes 
the fluid more resistant to breakup, allowing it to maintain rotational 
motion and develop a highly asymmetric contact surface over a wider range 
of vibration amplitudes and frequencies.
In contrast, less viscous fluids break more easily during rotation, 
so the growth rate can only be measured within a narrow range of lower 
vibration amplitudes and frequencies.

\subsection{Particle--particle and particle--substrate contacts}
\label{sec:contacts}
Determining the free energy (e.g., Helmholtz free energy) of the system
with respect to a reference state (e.g., equilibrium configuration of 
the droplet ``resting'' on the substrate) directly from the simulation
would require free energy expressions that describe the various
contributions beyond the ideal gas free energy. 
In the case of the droplet--substrate system of this study, the 
free energy terms are well established in the literature, and 
based on classical density functional theory, one can
calculate the various free energy contributions from the densities obtained
directly from the simulations.\cite{Theodorakis2022,Xu2007}
The contributions from interactions between particles, 
as well as between the particles and the substrate, are expected to play
the most prominent role. Evaluating these can be done through
an explicit expression for the potential, which in MDPD can be obtained
in an approximate form by considering both the attractive and
the repulsive contributions. However, this is in fact not necessary, 
since for systems where long-range interactions are negligible, the particle--particle and particle--substrate free-energy 
contributions are expected to be proportional to the normalized
number of particle--particle ($n_{\rm pp}$)
and particle--substrate ($n_{\rm ps}$) contacts for
any cutoff distance beyond the average distance of first neighbors,
according to the Stillinger criterion. \cite{Stillinger1963}
Here, the cutoff distances are chosen as $r_{d}=0.75$ for particle--particle interactions and $1.5\sigma_{\rm ws}=1.755$ for particle--substrate interactions.
Thus, these contacts are also expected to directly reflect the
particle--particle and particle--substrate energy contributions. 
For this reason, in the following, we attempt to characterize the state of
the system by using the number of contacts per particle.
Here, we show these values for each phase for the systems
discussed in Section~\ref{sec:rotation} 
for the sake of consistency.

\begin{figure}[htb!]
\includegraphics[width=0.99\columnwidth,trim=1.2cm 0cm 2.3cm 0.5cm,clip]{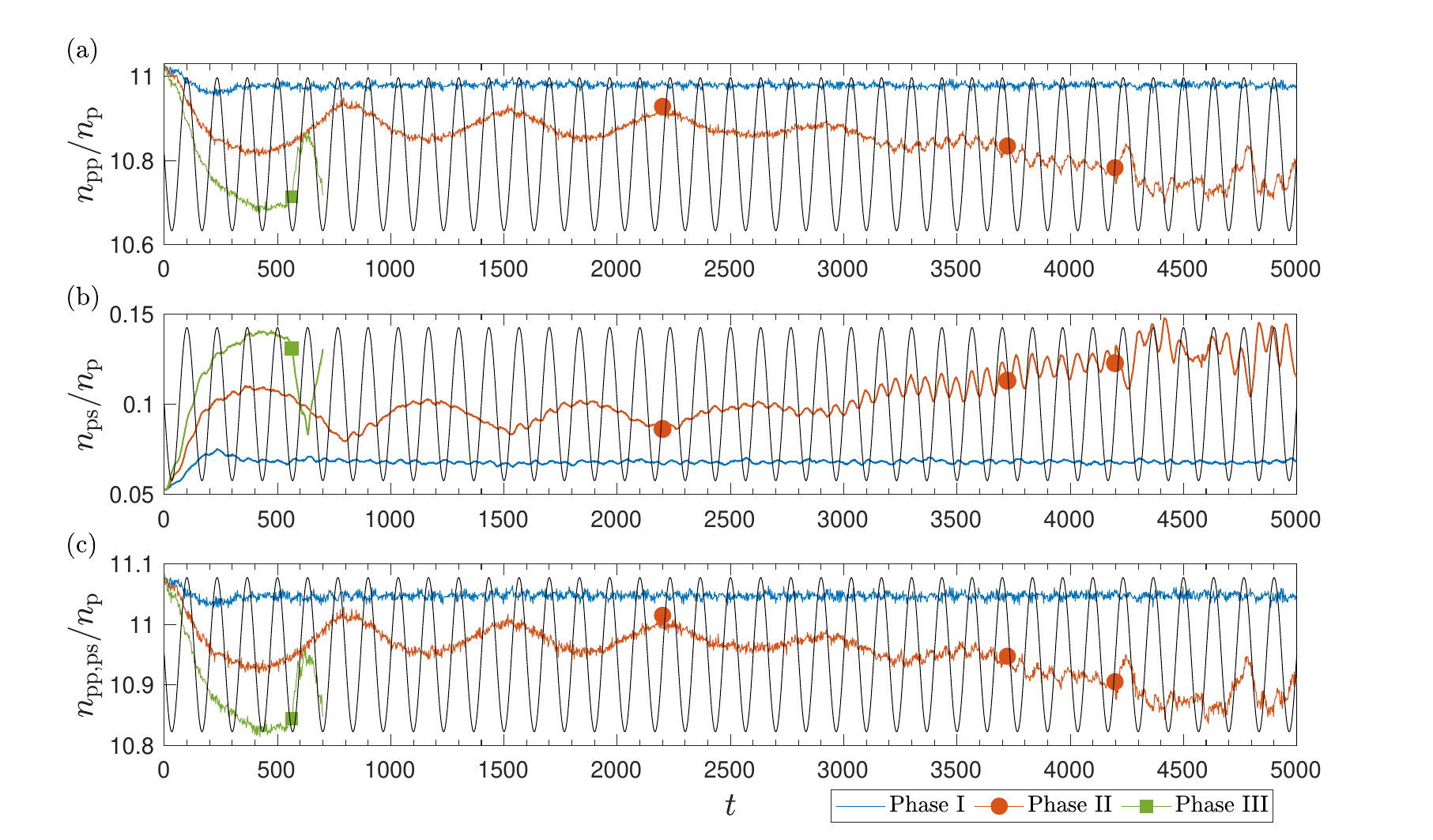}%[trim={left bottom right top},clip]
\caption{Number of contacts per particle over time for droplets
at $\rm{Oh}=0.1660$, 
equilibrium contact angle $\theta=90^{\circ}$, 
and particle number $N=20\times10^4$. 
The three phases are represented as follows: a blue line for phase I, 
a red line with dot markers for phase II, 
and a green line with a square marker for phase III. 
The dot markers in phase II correspond to the times 
$t=2200$, $3700$, and $4200$ as in Figure~\ref{fig:phase2_scalar}, 
while the square marker in phase III indicates the breakup time $t=562$, 
as in Figure~\ref{fig:phase3_scalar}. 
All three phases were generated using the same
substrate frequency $\omega_{\rm sub}=0.015\pi$, with different amplitudes $A_{\rm sub}$. 
The black curve shows the substrate velocity for reference, 
with its peaks normalized to the same height across all phases. 
(a) Number of particle--particle contacts ($n_{\rm pp}$) per particle,
where $n_{\rm p}$ is the instantaneous number of particles in the droplet at time $t$;
(b) number of particle--substrate contacts ($n_{\rm ps}$)
per particle; 
(c) total number of contacts ($n_{\rm pp,ps}=n_{\rm pp}+n_{\rm ps}$) per particle.}
\label{fig:combine_ncs}
\end{figure}

In the case of phase I oscillation (Figure~\ref{fig:combine_ncs}, blue lines),
we find that the droplet's contacts reach a steady-state equilibrium after
an initial stage of the oscillatory motion. While differences
over time in the number of contacts are rather small, with an initial contacts per particle
$n_{\rm pp}/n_{\rm p}\approx11.03$ 
for the example shown here (Figure~\ref{fig:combine_ncs}(a), blue line),
one can observe that $n_{\rm pp}/n_{\rm p}$ reaches a minimum of approximately $10.96$, shortly after 1.5 oscillation cycles ($t\approx200$) due to the initial extension of the droplet (see Figure~\ref{fig:combine_dia}(a)). The number of contacts per particle rebounds slightly to $n_{\rm pp}/n_{\rm p}\approx10.98$ after 3 oscillation cycles ($t\approx400$), following the re-contraction of the droplet.
This reduced number of contacts per particle suggests that the 
droplet is in a higher-energy state (due to the attractive nature of
the interactions between the particles, the energy is considered negative) compared to the initial, static, equilibrium
case before substrate vibration begins. 
After the initial transients, $n_{\rm pp}/n_{\rm p}$ continues to fluctuate around
its average value, and no clear correlation to the vibration period is
detected. The number of particle--substrate interactions
per particle, $n_{\rm ps}/n_{\rm p}$ (Figure~\ref{fig:combine_ncs}(b), blue line), shows an opposite trend compared to 
$n_{\rm pp}/n_{\rm p}$. It reaches a maximum value at the same time when $n_{\rm pp}/n_{\rm p}$ reaches its minimum, 
indicating an increased number of particle--substrate contacts. 
Taking both contributions into account together, 
namely $n_{\rm pp}/n_{\rm p}$ and $n_{\rm ps}/n_{\rm p}$, we find that the 
total number of contacts per particle during oscillation is lower than in the initial static case (Figure~\ref{fig:combine_ncs}(c), blue line).
This indicates that the
droplet overall exists in a less favorable energy state during oscillations compared to the static (no vibrations), equilibrium case of the
sessile droplet. 

Phase II exhibits a more characteristic behavior concerning the time
evolution of the contact numbers during oscillations 
(Figure~\ref{fig:combine_ncs}, red lines). 
Both $n_{\rm pp}/n_{\rm p}$ and $n_{\rm ps}/n_{\rm p}$ show an oscillatory pattern 
with a long period extending 
over multiple vibration cycles, where $n_{\rm pp}/n_{\rm p}$ 
reaches its maximum values 
(Figure~\ref{fig:combine_ncs}(a), red line) 
when $n_{\rm ps}/n_{\rm p}$ reaches its minima 
(Figure~\ref{fig:combine_ncs}(b), red line).
This oscillatory pattern aligns with the periodic contraction--extension of the
droplet up to the end of the elongation period at $t\approx2200$
(see Figure~\ref{fig:combine_dia}(b)), 
indicating that the periodic variations in both $n_{\rm pp}/n_{\rm p}$ 
and $n_{\rm ps}/n_{\rm p}$ are due to changes in the droplet's contact length.
The total number of particle--particle and particle--substrate contacts,
$n_{\rm pp,ps}/n_{\rm p}$ (Figure~\ref{fig:combine_ncs}(c), red line), follows the trend of $n_{\rm pp}/n_{\rm p}$, 
as fluctuations in $n_{\rm pp}/n_{\rm p}$ appear larger than those in
$n_{\rm ps}/n_{\rm p}$, presumably because the droplet contains more particles
in the bulk than at the solid--liquid interface.
After the initial oscillatory behavior of the ``free energy''
as tracked by the contact numbers, the rotational instability
begins to develop at $t\approx2200$.
From this point until the droplet breaks up at $t=4200$,
$n_{\rm pp}/n_{\rm p}$ shows a steady decrease from approximately
$10.93$ to $10.74$, 
while $n_{\rm ps}/n_{\rm p}$ 
shows a corresponding increase from approximately 
$0.086$ to $0.13$. 
The overall number of contacts per particle, $n_{\rm pp,ps}/n_{\rm p}$, also shows
a steady decrease, indicating that the system goes into
a less energetically favorable state, moving away from a metastable equilibrium.

Finally, phase III exhibits a rapid growth of the instability observed
in the case of phase II, as shown by the green lines in Figure~\ref{fig:combine_ncs}.
From the onset of substrate vibration,
the oscillating droplet follows a phase II-like behavior, 
but no oscillatory pattern is observed. 
This absence is due to the fact that, in phase III, 
the droplet's contact length rapidly increases until breakup occurs, 
without undergoing any periodic contraction--extension
(see Figure~\ref{fig:combine_dia}(c)). 
Moreover, the droplet reaches a deeper minimum in 
particle--particle contacts (Figure~\ref{fig:combine_ncs}(a), green line) and a higher maximum in 
particle--substrate contacts (Figure~\ref{fig:combine_ncs}(b), green line) than those observed in phase II.
Specifically, for the example shown here,
phase III reaches a minimum $n_{\rm pp}/n_{\rm p}\approx10.68$
and a maximum $n_{\rm ps}/n_{\rm p}\approx0.14$ 
before the breakup at $t=562$.
It is also interesting to observe that once the breakup occurs, 
the largest child droplet formed on either side of the parent droplet attains a
slightly more energetically favorable state, 
as indicated by a higher overall number of contacts per particle, $n_{\rm pp,ps}/n_{\rm p}$ (Figure~\ref{fig:combine_ncs}(c), green line).

In summary, the results in Figure~\ref{fig:combine_ncs} 
indicate that throughout the simulation experiments,
for droplets with the same number of particles and equilibrium 
contact angle, phase I oscillation 
always exhibits the highest number of contacts per particle compared to the other
phases, that is, the most stable mode energetically rather pointing to a steady state. In general, one may argue that phase II exhibits a higher number of overall contacts per particle compared to phase III, indicating
metastable energy minima up to the droplet breakup,
while phase III is an out-of-equilibrium state of the system.

\section{Conclusions}
\label{conclusions}

In this study, we performed MDPD simulations across a range of different liquids, wettable substrates, and substrate vibration modes to investigate 
droplet oscillations induced by horizontal substrate-vibrations. The main findings of our study are summarized as follows:

\renewcommand{\labelitemi}{$\Diamond$}
\begin{itemize}
    \item We confirmed the scaling for the 
    fundamental frequency of droplets containing more than $50\times10^3$
    particles, specifically for droplets on less wettable substrates. As seen in Figure~\ref{fig:freqz_dcomz}, 
    sessile droplets with an equilibrium contact angle of $\gtrsim110^\circ$ behave like freely
    suspended drops regarding their fundamental frequency.
    Moreover, we demonstrated that this is
    strongly influenced by the presence of the substrate --- a
    more wettable one leads to
    significant deviations from the Rayleigh theoretical
    predictions.\cite{Rayleigh1945}
    \item We found that the capillary number Ca provides 
    a suitable measure for categorizing the various
    states of the system, namely phases I, II, and 
    III. Our analysis also allowed us to determine
    the critical Ca values for the phase transitions from I to II and from II to III for various contact angles.
    \item We identified the key characteristics
    of phase II oscillations and introduced suitable
    parameters for characterizing the emergence
    of phase II, namely the angular momentum and vorticity
    of the droplet. Based on these parameters, 
    we also determined the growth rate of the instability
    as a function of the Ohnesorge number in phase II oscillations.
    \item In the absence of a potential form in MDPD simulations, 
    we demonstrated that monitoring the number
    of particle--particle and particle--substrate contacts
    provides a way of gaining insights into the state of the droplet. Based on these properties, we discussed the dynamic behavior and energy characteristics of each oscillation phase and associated those with
    the conformational changes of the droplets
    during oscillation. 
\end{itemize}

We anticipate that the above insights reflect the fundamental knowledge
required for understanding droplet oscillation phenomena on horizontally
vibrating substrates, holding much relevance for applications, such as
lab-on-a-chip devices, microreactors, single-cell sorting, and
drug delivery.

\begin{acknowledgments}
This research has been supported by the National 
Science Centre, Poland, under
Grant No.\ 2019/34/E/ST3/00232. 
We gratefully acknowledge the Polish high-performance computing 
infrastructure PLGrid (HPC Center: ACK Cyfronet AGH) 
for providing computer facilities and support within 
computational Grant No. PLG/2024/017543. T.B. and M.K. acknowledge the
support from Warsaw University of Technology within the
Excellence Initiative: Research University (IDUB) program.
\end{acknowledgments}

%\nocite{*}
\bibliography{aipsamp}% Produces the bibliography via BibTeX.

\newpage
%\subsection{Supplementary Material 1}
%These are the state diagrams for medium and smaller sized
%droplets.

\end{document}